\documentclass{article}

\usepackage[nonatbib,preprint]{neurips_2021}

\usepackage[utf8]{inputenc} 
\usepackage[T1]{fontenc}    
\usepackage{hyperref}       
\usepackage{url}            
\usepackage{xurl}
\usepackage{hyperref} 
\usepackage{booktabs}       
\usepackage{amsfonts}       
\usepackage{nicefrac}       
\usepackage{microtype}      
\usepackage{xcolor}         
\usepackage{graphicx}
\usepackage{lineno,amsfonts,amsmath}
\usepackage[most]{tcolorbox}
\usepackage{authblk}

\newcounter{manuscriptbox}

\usepackage{algpseudocode,algorithm}

\algrenewcommand\algorithmicrequire{\textbf{Precondition:}}  
\algrenewcommand\algorithmicensure{\textbf{Postcondition:}}

\title{Anthropomorphism in the age of Large Language Models: An overview of potential risks and mitigations}

\author[1]{Ismael T. Freire*}
\author[1,2]{Marceau Nahon*}         
\author[1]{Maud van Lier}
\author[2]{Katie Evans}
\author[1,2,3]{Hélie Bazin}
\author[4,5]{Michele Farisco}
\author[4]{Kathinka Evers}
\author[1]{Raja Chatila}
\author[1]{Mehdi Khamassi}

\affil[1]{Institute of Intelligent Systems and Robotics, CNRS, Sorbonne University, Paris, France}
\affil[2]{Sciences, Norms, Democracy, CNRS, Sorbonne University, Paris, France}
\affil[3]{Sorbonne Center for Artificial Intelligence, Sorbonne University, Paris, France}
\affil[4]{Centre for Research Ethics and Bioethics, Uppsala University, Uppsala, Sweden}
\affil[5]{Bioethics Unit, Biogem, Biology and Molecular Genetics Research Institute, Ariano Irpino, Italy.}

\affil[*]{\rule{0pt}{2em}\normalfont\upshape Corresponding co-first authors: 
  \texttt{\{ismael.freire,marceau.nahon\}@isir.upmc.fr}}

\begin{document}

\maketitle

\begin{abstract}
Large Language Models (LLMs) and more broadly Artificial Intelligence (AI) systems are often described and understood in human-like terms, a phenomenon known as \emph{anthropomorphism}. This paper provides a synthesis of recent literature on anthropomorphism in AI, covering theoretical frameworks, the role of language in framing AI as human-like, the various risks of anthropomorphizing machines, and strategies to mitigate these issues. After examining why we tend to anthropomorphize AI systems and whether we are right to do so, we highlight the impact of linguistic framing on anthropomorphism. Then, we introduce a conceptual taxonomy of risks associated with AI anthropomorphism. This taxonomy groups twenty-one concerns within five analytical categories: epistemic, affective, human agency, normative, and societal and institutional risks. Finally, we relate these concerns to proposed interventions in design, communication, education, and governance. We argue that a better understanding of AI systems requires concepts and theories grounded in their organization and demonstrated capacities. The linguistic shaping of anthropomorphic perceptions should form part of this scientific effort, since our descriptions influence both how these systems are understood and the roles we allow them to occupy in society.
\end{abstract}

\section{Introduction}
\label{sec:1_intro}
\emph{Anthropomorphism} is a natural human tendency to project human-like qualities onto non-human entities \cite{stebbins1993anthropomorphism}. In the context of artificial intelligence (AI), this often means treating AI systems as if they have minds, intentions, or emotions similar to our own. What is particularly different from the anthropomorphism of other non-human entities is that on top of treating AI systems as if they have human-like capacities,  we also act on the basis of the suggestions and decisions that they produce, appraising these in ways similar to how we would evaluate those of human counterparts. As AI systems become more sophisticated in what they can do \cite{salles2020anthropomorphism, akbulut2024all}, their practical influence over our lives increases. For example, advanced conversational agents can by now produce fluent and contextually appropriate language, leading people to find it "increasingly hard to resist" ascribing human qualities to them \cite{Shanahan2024Talking}, treating them effectually as they would a friend or a trusted expert. This is deeply problematic as long as it is still a question whether AI systems actually possess the capacities that we attribute to them \cite{Birhane2023Science,Goddu2024LLMs}. In some contexts, for instance, anthropomorphism can operate as a cognitive illusion that compromises epistemic autonomy by biasing users to interpret system outputs through an unwarranted model of human-like mentality. From this perspective, anthropomorphism is an interpretative stance with epistemic and ethical implications, and therefore, a conceptual analysis is a prerequisite for normative debate \cite{salles2020anthropomorphism}.

Anthropomorphism of AI systems can stem from superficial resemblances (e.g., natural language use, affective tone, giving the system eyes), the ability to perform tasks that we usually perform, the context in which a system is placed, and from the language we use to describe AI systems. Even experts themselves are prone to using intentional language for AI systems, inadvertently "begging the question" by implying the very understanding or agency that is in dispute \cite{Shanahan2024Talking}. Moreover, the anthropomorphic habit of thought is reinforced by decades of science fiction and cultural narratives that portray AI systems as near-human or even super-human intelligences. Yet, human-like performance by an AI system does not necessarily equate to a human-like mind \cite{Shanahan2023Role,Shanahan2024Talking}.

This article offers a narrative and conceptual review rather than an exhaustive or systematic survey. It focuses primarily on large language model-based conversational systems, while drawing on broader research concerning AI, robotics, human-computer interaction, and mind perception where this helps clarify the relevant mechanisms and concepts. Section \ref{sec:2_why} examines the cognitive, motivational, cultural, and linguistic factors that make anthropomorphic interpretations appealing, including the distinctive role of natural-language interaction. Section \ref{sec:3_theoretical_frameworks} evaluates whether the technical capacities of contemporary AI systems warrant attributions of agency, intentionality, consciousness, knowledge, understanding, reasoning, or moral competence. Section \ref{sec:4_language_metaphors} considers how terminology and metaphor mediate the transition from observed performance to claims about underlying capacities. Section \ref{sec:5_risks} organizes the risks of anthropomorphizing AI systems into five overlapping domains, and Section \ref{sec:6_mitigation_strategies} reviews proposed responses to mitigate these risks in design, communication, education, institutional practice, and governance. Finally, Section \ref{sec:7_discussion} summarizes the resulting case for calibrated, capacity-specific descriptions of contemporary AI systems.

\section{Why Do We Anthropomorphize?}
\label{sec:2_why}

\subsection{Cognitive Sciences of Anthropomorphism}
\label{subsec:2_1_cognitive_sciences}

Human experience is inherently self-referential: perception, memory, valuation, and action are organized from the organism's own embodied and affective standpoint. Changeux and Connes argue that human access to reality is mediated by brain-generated models constrained by the physical organization of the brain \cite{changeux1995conversations}, while Evers describes the resulting epistemic finitude as the condition of being, in a sense, "prisoners of our brains" \cite{evers2009neuroethique}. A related perspective can be found in von Uexküll's notion of the \emph{Umwelt}, according to which organisms encounter their surroundings through species-specific perceptual and action capacities that structure what becomes salient and meaningful to them \cite{uexkull2021foray}. Human cognition is also self-projective, allowing individuals to imaginatively place themselves in alternative situations and to represent the perspectives of others \cite{buckner2007self}. Together, these characteristics make one's own mental life and human mode of experience particularly accessible reference points when interpreting other entities. In situations involving ambiguity, complexity, or social salience, familiar human categories such as beliefs, intentions, desires, and emotions can provide readily available models for making sense of non-human behaviour \cite{epley2007seeing, epley2008we, epley2018mind}.

These cognitive tendencies can also be considered from an evolutionary perspective. Anthropomorphism has been argued to draw on capacities that were adaptive in the social and ecological environments in which human cognition evolved \cite{guthrie2013anthropomorphism, beer2014perspectives, dacey2017anthropomorphism}. For a highly social primate species, attributing mental states similar to its own to others is a core competence since it makes it possible to anticipate what the other might do, and adjust one's own behaviour accordingly. This resonates with the \emph{social brain hypothesis}, in which the evolution of primate brains is linked to the cognitive demands of managing complex social relations \cite{dunbar1998social}. Similarly, Tomasello and colleagues argue that human cognition is distinctively shaped by \emph{shared intentionality}, intention-reading, cooperation, and cultural learning \cite{tomasello2005understanding}. Attributing similar kinds of beliefs, emotions, intentions, and goals to other humans and animals (and in some circumstances also to inanimate objects) would have been useful to predict behaviour, coordinate action, detect threats, and adapt to social environments. Colin Beer posits that the attribution of mental states is permitted by what he refers to as a \emph{social tool kit} developed during evolution \cite{beer2014perspectives}. This tool kit can explain why children use a lexicon of intentionality when talking about clouds \cite{laurendau1962causal}, and why they have a strong tendency to attribute intelligence, desires, beliefs and intentions to non-human animals \cite{searle1999mind, myers2002animals, serpell2002anthropomorphism, goldman2024children}. It could also explain why we have a tendency to perceive human traits in non-human objects (such as faces in the clouds), a phenomenon called \emph{pareidolia}.

The cognitive sciences of religion also offer a relevant account of the cognitive mechanisms behind anthropomorphism, as well as that of animism (see Box \ref{box:terminological_distinctions}). Justin Barrett, for instance, in trying to explain the cognitive origins of religion, introduces the \emph{Hyperactive Agency Detection Device} (HADD) \cite{barrett2000exploring, barrett2007cognitive}. He argues that our mechanisms to ascribe agency are hyperactive, leading to the overattribution of agency and intentionality to events and entities that do not possess it, which leads to anthropomorphism. Following Stewart Guthrie \cite{guthrie1980cognitive, guthrie1995faces}, Barrett argues that such an overdetection of intentional agents could confer an evolutionary advantage even when it is misplaced, as false-positives were far less costly than false-negatives: mistaking a wind-blown bush for a predator is less costly than failing to detect a predator concealed within it  \cite{barrett2007cognitive}. Lisdorf’s \emph{Hyperactive Intentionality Detection Device} (HIDD) refines this account by stressing our tendency to project intention and purpose, not just agency, when dealing with uncertainty over ambiguous events \cite{lisdorf2007s}. 

These individual cognitive dispositions are also filtered and amplified by culture. According to Boyer, anthropomorphic representations are culturally successful because they are easy to remember and to transmit, and therefore to spread, because they invite people to explain the functioning of non-human systems through familiar mentalistic categories such as beliefs, desires, intentions, or emotions  \cite{boyer1996makes, mithen1996anthropomorphism}. Consequently, cultural transmission could play a role in propagating particular anthropomorphic framings, by making them socially available. Several studies have for example identified cross-cultural variations in the attribution of minds to non-biological entities \cite{papadopoulos2018influence}. Spatola and colleagues propose a tripartite model describing anthropomorphism as the result of three independent processes: \emph{mentalisation} (attributing cognition), \emph{humanisation} (attributing emotion) and \emph{animism} \cite{spatola2022different}. Participants are shown examples of human–robot interaction and asked to evaluate the plausibility of mental, human or spiritual interpretations of the scene. Spatola and colleagues found that the individual animistic beliefs of participants influence all three levels. Furthermore, they discovered that the impact of animism varies depending on the participants' origin: the animistic beliefs of East Asian participants positively influence mentalisation, but have no effect on humanisation; the opposite is true for Western participants. This suggests that, although personal beliefs (animism) impact overall anthropomorphism, culture influences how the anthropomorphism process is achieved. 

Anthropomorphism might further be reinforced by the very grammatical structure of Western languages. In Chomsky's \emph{Government and Binding theory} \cite{chomsky1981lectures}, for instance, predicates assign theta-roles to their arguments that specify how each argument relates to the predicate. Verbs such as "say" can encourage readers to interpret their grammatical subjects as intentional speakers. In the sentence "ChatGPT says that Paris is the capital of France," "ChatGPT" is attributed an agent-role by the verb "says". Thus, the use of active verbs may be interpreted as presupposing that there is an agent that is intentionally acting. Even though this might not constitute a proper cause of anthropomorphism, linguistic agency could enhance it. This last point is especially relevant for AI systems: once a Large Language Model (LLM) is represented as a speaker, assistant, companion, or agent, users can effortlessly import a whole repertoire of human social expectations into the interaction.

\refstepcounter{manuscriptbox}
\label{box:terminological_distinctions}
\begin{tcolorbox}[
    title={Box~\themanuscriptbox. Terminological distinctions: anthropomorphism, animism, and related notions},
    colback=gray!5,
    colframe=gray!60,
    fonttitle=\bfseries,
    breakable
]

Several related concepts recur in discussions of artificial intelligence, religion, and human-machine interaction. Some originate in existing psychological, philosophical, and social-scientific literatures, while others are introduced in this paper to capture distinctions that become particularly relevant for contemporary AI. We use them as follows.

\textbf{Anthropomorphism} refers to the attribution of human-like properties to non-human entities, including human-like mental states, emotions, intentions, motivations, agency, or social capacities \cite{epley2007seeing,guthrie1995faces}. In the context of AI, anthropomorphism is primarily an interpretative act performed by users, observers, designers, or institutions: an artificial system is treated as if it possessed human-like understanding, concern, autonomy, or personhood, whether or not such capacities are warranted by its architecture or behaviour.

\textbf{Mind perception} is the broader psychological process by which observers attribute mental capacities to entities. These capacities are often organized along dimensions such as \emph{agency} (planning, self-control, communication, moral action) and \emph{experience} (feeling, pain, pleasure, hunger, or consciousness) \cite{gray2007dimensions,waytz2010sees}. This distinction is especially important for AI ethics, because attributing competence or agency to a system does not automatically imply that the system has subjective experience, welfare, or moral patienthood.

\textbf{Animism} refers to the attribution of life, soul, spirit, vitality, or personhood to non-human beings, objects, places, or natural phenomena \cite{tylor1871primitive,boyer1996makes}. Animism overlaps with anthropomorphism when the attributed agency is specifically human-like, but the two are not identical. One may treat a river, mountain, forest, animal, or machine as animated, alive, or spiritually significant without necessarily attributing to it a human psychology.

\textbf{Sociomorphing} refers to the perception of actual non-human social capacities \cite{seibt2020sociomorphing}. Research on sociomorphing argues that sociality is not inherently human, stating that people respond socially to entities they do not anthropomorphize.

\textbf{Mechanomorphism} refers to attribution of machine-like properties to non-machine entities \cite{caporael1986anthropomorphism,karlsson2012anthropomorphism}. We argue that seeing the human in the machine can lead to seeing the machine in the human.

\textbf{Pareidolia} is a perceptual phenomenon in which ambiguous or random stimuli are experienced as meaningful patterns, most prominently as faces \cite{palmer2020face,wardle2020rapid}. Pareidolia may provide a perceptual entry point for anthropomorphism (for example, when a face-like arrangement on a robot or interface elicits social expectations), but it does not itself entail the attribution of mental states, intentions, or moral standing.

\textbf{Linguistic pareidolia} refers to the tendency to perceive evidence of an underlying mind, understanding, or experiencing subject in linguistic patterns that resemble human communication. By analogy with perceptual pareidolia, in which meaningful forms such as faces are perceived in ambiguous visual stimuli, linguistic pareidolia describes how coherent, responsive, or affectively expressive language can evoke the impression of a mind behind the words.

The \textbf{ELIZA effect} names a specific form of AI-related anthropomorphism in which users read more understanding, empathy, or intelligence into a computer program than its mechanisms justify \cite{weizenbaum1966eliza,weizenbaum1977computer}. Contemporary large language models intensify this risk because fluent dialogue, memory-like personalization, and socially responsive language can make statistical or computational processes appear as understanding, intention, or care.

\textbf{Anthropomorphic design} or \emph{anthropomimesis} refers to the deliberate introduction of human-like cues into artificial systems, such as names, voices, faces, conversational turn-taking, emotional language, or first-person self-reference. This should be distinguished from anthropomorphism itself: anthropomimesis concerns how a system is designed or presented, whereas anthropomorphism concerns how it is interpreted \cite{nass2000machines}.

\textbf{Agentic tool} refers to an artificial system capable of pursuing specified goals and acting with some degree of operational autonomy, while its objectives, constraints, and conditions of deployment remain grounded in human design and use. The concept acknowledges genuine capacities for autonomous action without presupposing human-like intentions, motivations, or moral responsibility. It distinguishes functional or operational agency from stronger forms of agency involving intentionality or responsibility.
\end{tcolorbox}

\subsection{The Specificity of LLMs: How not to Anthropomorphize Machines that Talk?}
\label{subsec:2_2_llm_specificity}
Long before the emergence of LLMs, linguistic performance was regarded as important behavioural evidence of thought or intelligence. Descartes maintained that the context-appropriate production of meaningful words or other signs was the only reliable external indication of thought. Although machines might produce particular utterances in response to certain stimuli, he argued that neither machines nor non-human animals could arrange meaningful signs appropriately across the indefinitely varied circumstances to which human beings can respond \cite{descartes1953lettre_newcastle,descartes1953lettre_morus}. For Descartes, this generality was evidence of reason that could not be explained by just the successful execution of a fixed mechanical disposition.

Linguistically mediated interaction later occupied a different role in Turing's discussion of machine intelligence \cite{turing2007computing}. Instead of defining what it means for a machine to think, Turing replaced the question "Can machines think?" with the more operational question of whether a digital computer could perform successfully in the \emph{imitation game}. More concretely, in this game, an interrogator attempts to distinguish a machine from a human participant through text-based exchanges, where language provides the interface through which a wide range of capacities can be probed. Importantly, neither the test was restricted to linguistic performance in a narrow sense nor did Turing establish indistinguishability from a human as a logically sufficient condition for thought. His proposal shifts the discussion from an elusive definition of thinking to an empirically assessable comparison of human and machine performance.

Despite the differences between their projects, both Descartes and Turing gave linguistic interaction an important evidential role in judgments about the presence of thought or intelligence. Until recently, machines generally failed to sustain sufficiently varied and context-sensitive linguistic exchanges for this comparison to become pressing. 

LLMs disrupt this practical association between linguistic and cognitive capacities. They are among the first technological artifacts capable of generating extended, novel, and contextually appropriate natural-language outputs across a wide range of domains. Because language is ordinarily produced by persons and interpreted as an expression of their beliefs, intentions, emotions, and understanding, such outputs strongly invite anthropomorphic interpretation. Knowing that an interlocutor is a computational system may be insufficient to prevent users from responding to it as though it possessed a human-like mind. We propose the term \textbf{linguistic pareidolia} for this phenomenon: the perception of human-like mentality on the basis of linguistic patterns that resemble the observable products of human thought, even when those patterns do not provide sufficient evidence for the attributed psychological capacities. The analogy with perceptual pareidolia is structural rather than sensory. In visual pareidolia, a configuration of features triggers the perception of a face despite not having been produced by a face; in linguistic pareidolia, patterns such as coherence, self-reference, emotional expression, or apparent responsiveness trigger the perception of an experiencing and understanding subject. The term identifies a potential mismatch between the strength of the anthropomorphic impression and the evidence available for the underlying capacities attributed to the system. However, this does not imply that every psychological attribution to an artificial system is necessarily mistaken. In the following section, we examine whether the linguistic capability of current LLMs warrants such attributions by considering their internal organization and the mechanisms through which their outputs are generated.

\section{Theoretical Frameworks: Are We Right to Anthropomorphize AI  Systems?}\label{sec:3_theoretical_frameworks}

Assessing the justification of anthropomorphism requires separating observable performance from claims about the capacities underlying it. Contemporary AI systems can display capacities that support functional or "as-if" descriptions involving goals, memory, inference, or social responsiveness. Whether these descriptions warrant stronger attributions of intentionality, consciousness, understanding, or moral competence is a separate and capacity-specific question. The following subsections review the principal positions and evidence relevant to that distinction.

\subsection{Agency and Intentionality}
\label{subsec:3_1_agency_intentionality_free_will}
Empirical work in social perception suggests that human observers can readily attribute agency and intentionality even to minimal non-human stimuli, provided that their behaviour exhibits appropriate cues. In the classic Heider-Simmel animation, participants interpreted the movements of simple geometric figures (such as triangles and circles) as a meaningful social scene involving pursuit, conflict, fear, and escape \cite{heider1944experimental}. Later work on perceptual animacy has extended this finding, showing that simple motion patterns can elicit impressions of animacy, goal-directedness, and intentional interaction \cite{scholl2000perceptual,parovel2023perceiving}. AI systems are not exceptional in inviting intentional descriptions: they constitute a particularly salient contemporary case in which complex, goal-directed, linguistic, or socially responsive behaviour encourages observers to adopt an intentional stance \cite{nass2000machines}.

Conversational systems provide a particularly clear case of this tendency. The so-called \emph{ELIZA effect} describes the propensity of users to attribute understanding, empathy, or intentionality to a computer program on the basis of relatively superficial linguistic cues, as already observed in reactions to Weizenbaum's early chatbot ELIZA \cite{weizenbaum1966eliza,weizenbaum1977computer}. This effect is important because it shows that intentional attributions in human-computer interaction can arise from the observer's interpretive dispositions rather than from the intrinsic capacities of the system. Cappelen \& Dever argue that the impression of linguistic agency quickly disappears when interacting with ELIZA, whereas it remains unchallenged when interacting with LLMs \cite{cappelen2025going}. On Cappelen and Dever's account, sustained, seemingly intentional dialogue would distinguish interactions with LLMs from the superficial linguistic cues associated with ELIZA. This interpretation, however, faces two challenges: Firstly, accuracy can decline during lengthy, multi-turn conversations \cite{maharana2024evaluating,laban2026llms}; and secondly, even where consistency is maintained, it does not establish that users' attributions are warranted. Improved performance could instead strengthen the ELIZA effect.

According to Dennett, the activities of some systems are most effectively predicted and explained from what he calls the \emph{intentional stance}: a strategy in which one treats the system as if it had beliefs, desires, goals, or other intentional states \cite{Dennett1971, dennett1988intentional}. This strategy is not restricted to artificial intelligence. Dennett applies it more broadly to humans, non-human animals, machines, and even hypothetical alien creatures, insofar as their behaviour can be usefully interpreted as rationally directed by information and goals. Dennett takes the example of a chess program: in many contexts, it is more practical to predict the program’s next move by saying that it "wants" to protect its queen or "believes" that a bishop is threatened than by describing its behaviour at the physical or implementation level.  Importantly, the intentional stance is a pragmatic and explanatory strategy without any metaphysical implication: adopting it does not assume anything about the real capacity or nature of the thing considered. 

Lavazza and colleagues make a related distinction when arguing that LLMs may satisfy some functional or behavioural criteria relevant to legal responsibility while nevertheless remaining dependent on external inputs and lacking intrinsic volition \cite{lavazza2026minds}. Yet, in the specific case of stand-alone LLMs, even the appeal to behavioural legal criteria must be treated with caution: linguistic output is not equivalent to autonomous behaviour in the richer sense of embodied, situated, and causally controlled action. At most, such accounts establish a functional or as-if notion of agency, not the presence of intentionality in the stronger sense relevant to free will.

A similar ambiguity appears in debates about communicative intention. Focusing specifically on LLMs, some authors argue that these systems lack communicative intentions. Montemayor, for example, argues that LLMs (specifically GPT-3 at the time) have no communicative intentions because they lack communicative goals or motivations and the capacity to participate in jointly purposeful interaction \cite{montemayor2021language}. Attah has recently challenged this conclusion, arguing that LLMs may possess communicative intentions in a weak sense, understood in terms of "degrees of freedom within a system of choices" \cite{attah2025language}. According to Attah, the difference between LLMs and humans would mainly lie in the absence of \emph{metalinguistic agency}. However, this proposal leaves open whether a sufficiently rich space of possible outputs is enough to support an attribution of communicative intention. Features such as independently generated goals, motivations, metalinguistic control, and participation in situated joint activity remain relevant to how such attributions are assessed, even if their precise role is philosophically contested. Current LLMs may invite intentional descriptions and may display increasingly sophisticated intentional-like behaviour, but this should not be confused with evidence that they possess intentionality as such. Claims about artificial intentionality require independent theoretical and empirical support rather than following directly from the adoption of intentional vocabulary.

Regarding the philosophical meaning of "intentionality", i.e., referring to something, a similar point applies to Grindrod who applies Millikan's theory of \emph{linguistic intentionality} to LLMs \cite{grindrod2024large}. For Millikan, linguistic forms have specific functions. When the function of a form is performed using a mapping between the form and some external state-of-affairs, the form bears intentionality as it has truth-conditions \cite{millikanLanguageBiologicalModel2005}. For Grindrod, LLMs produce words by exploiting a mapping between their internal states and an external state-of-affairs, and as such bear linguistic intentionality. However, even if Millikan's theory is true and successfully applies to LLMs, it remains neutral as to whether LLMs perform the mapping intentionally and it thus cannot justify the attribution of mental properties to LLMs. All in all, these considerations suggest there is no strong evidence that LLMs and other AI systems genuinely possess the intentionality or the forms of agency that we attribute to them in certain contexts.

\subsection{Consciousness}
\label{subsec:3_2_consciousness}
The question of consciousness in artificial systems is particularly difficult because, among other things, at the theoretical level there is currently no complete or universally accepted theory of consciousness. Claims about whether AI systems could be conscious depend, explicitly or implicitly, on background assumptions about the definition and the nature of consciousness itself. This does not mean that nothing is known about consciousness, however. On the contrary, cognitive neuroscience, comparative psychology, and philosophy of mind have produced a substantial body of work concerning consciousness in biological organisms, as well as a plurality of theoretical frameworks for describing and possibly explaining it. Kuhn's extensive review of theories of consciousness provides an exhaustive depiction of a heterogeneous landscape, ranging from physicalist and biological accounts to computational, functionalist, panpsychist and non-physicalist positions \cite{kuhn2024landscape}.

Computationalist and functionalist accounts are generally more open to the possibility of artificial consciousness. On these views, consciousness is substrate-independent: it depends on, or is constituted by, an appropriate functional or computational organization rather than by the particular material (either biological or artificial) in which that organization is realized. If the relevant, necessary and sufficient organizations exist in a non-biological system, artificial consciousness is possible in principle \cite{chalmers1997conscious,chalmers2023could}. This does not entail that current AI systems are conscious. Chalmers, for example, argues that current LLMs face important obstacles, including the absence of recurrent processing, global workspace-like organization, and unified agency. Even so, he takes the possibility seriously that future systems could overcome such limitations \cite{chalmers2023could}. Similarly, Butlin and colleagues propose a theory-driven framework for identifying indicators of consciousness in AI systems, drawing on \emph{recurrent processing theory}, \emph{global workspace theory}, \emph{higher-order theories}, \emph{predictive processing}, and \emph{attention schema theory} \cite{butlin2023consciousness,butlin2025identifying}. Their conclusion is cautious: current AI systems do not appear to be conscious, but future systems might be designed to satisfy some of the relevant indicators.

Other accounts place stronger conditions on the possibility of artificial consciousness. \emph{Biological naturalism}, in Searle's sense, treats consciousness as a biological phenomenon caused by and realized in physical mechanisms with the relevant causal powers, instead of as an abstract computation considered independently of its material realization \cite{searle2017biological}. Related embodied and organismic approaches assign greater significance to metabolism, self-maintenance, and the distinctive organization of living systems and nervous systems \cite{godfrey2016mind,godfrey2023nervous,thompson2010mind,thompson2022could,seth2021being}. However, these features do not necessarily have to be non-functional. The key is that these approaches question whether the properties relevant to consciousness can be specified entirely at an abstract functional level while disregarding how they are physically and biologically realized. In other words, we do not know which level of detail of the physical implementation of consciousness is actually necessary for it \cite{cao2022multiple, block2025can, changeux2026global}.

From this perspective, Seth argues that reproducing the behavioural or computational functions associated with consciousness would not, by itself, establish the presence of conscious experience  \cite{Seth2025Conscious}. Present-day LLMs can produce fluent reports of feeling, understanding, or self-awareness. Such a linguistic expression may simulate the outward signs commonly associated with consciousness, but it does not demonstrate that there is an experiencing subject or some sort of inner life behind those reports. Similar caution is found in neuro-scientific critiques of artificial consciousness, which emphasize the absence in current AI systems of key structural, evolutionary, embodied, and organismic features associated with consciousness in biological agents \cite{aru2023feasibility,farisco2024artificial,chen2025exploring}.

The current balance of expert opinion also supports caution. In the 2020 PhilPapers survey, only a small minority of professional philosophers accepted or leaned towards the view that current AI systems are conscious (3.39\%), whereas a large majority rejected or leaned against it (82.42\%) \cite{bourget2023philosophers}. A recent survey of AI researchers likewise found low median credence that AI systems with subjective experience already existed in 2024 (1\%), while reporting substantially greater uncertainty about future systems (25\% by 2034 and 70\% by 2100) \cite{dreksler2025subjective}. Thus, the outcomes of these surveys call for skepticism about consciousness in present systems alongside uncertainty about future systems. This distinction is nonetheless important: rejecting consciousness in current LLMs does not require rejecting artificial consciousness in principle.

In anthropomorphism, however, the crucial issue is that people attribute human-like consciousness to AI systems, whether or not they are actually conscious. Empirical work shows that many users are willing to attribute some degree of phenomenal consciousness to LLMs, and that such attributions increase with familiarity and usage frequency \cite{colombatto2024folk}. This mirrors the broader pattern discussed above for agency and intentionality, where anthropomorphic attribution can occur independently of the system's actual capacities. For this reason, Shanahan and colleagues suggest understanding LLM dialogue in terms of \emph{role-play}, as a way of describing their human-like linguistic performance without ascribing to them the human properties they lack \cite{Shanahan2023Role}. Interpreting an AI system's dialogue as \emph{pretend play}, rather than as genuine expression grounded in an inner self, helps preserve the distinction between linguistic performance and subjective agency. The system generates outputs through learned computational mappings conditioned on the interaction context, as first-person or affective wording does not establish an experiencing subject. This distinction is ethically important because competent human-like linguistic behaviour should not be treated as sufficient evidence of consciousness.

At the same time, resisting anthropomorphic interpretations does not require ruling out the possibility of artificial minds or consciousness in principle. Related work on artificial consciousness has similarly emphasized that whether artificial systems could instantiate dimensions or forms of consciousness remains an open theoretical and empirical question \cite{Evers2025Preliminaries}. Whether artificial minds are theoretically or technically possible is out of the scope of this paper, but even if they were, there is no reason to assume that their organization or phenomenology would necessarily resemble our own. Sloman's notion of a \emph{space of possible minds} similarly invites consideration of forms of mentality that may differ substantially from familiar human cases \cite{sloman1984structure}. For the present review, we distinguish this possibility from the effects of systems that appear conscious to their users. Seth, for instance, makes a similar distinction between the prospects of conscious AI systems and the ethical and societal implications of \emph{seemingly conscious AI systems}, emphasizing that the latter could arise independently of whether consciousness is genuinely instantiated \cite{seth2026stuff}. This contrast is especially relevant to the anthropomorphism debate surrounding AI, since human-like linguistic and social cues can shape users' beliefs and responses even in the absence of reliable evidence about the systems' subjective experience. Because of this, we should pay attention to how anthropomorphic attributions arise, how perceptions of consciousness shape human responses, and how possible forms of artificial mentality should be investigated on their own terms. So in short, while the broader possibility of distinct forms of artificial consciousness remains an open question, for current LLMs, human-like dialogue alone does not provide sufficient grounds for attributing human-like subjective experience to these systems.

\subsection{Knowledge, Understanding and Reasoning}
\label{subsec:3_3_knowledge_understanding_reasoning}

Researchers have also studied the notions of knowledge and understanding in AI systems. Yildirim and Paul introduce the concept of \emph{instrumental knowledge} for LLMs: these models have knowledge in the limited sense of statistical connections enabling useful predictions, but not in the human sense of grounded, conceptual understanding \cite{Yildirim2024Instrumental}. By coining new terms like \emph{instrumental knowledge}, they aim to avoid conflating AI systems' abilities with human cognitive states. In response, other cognitive scientists have gone further to argue that LLMs "don’t know anything" in the human sense at all \cite{Goddu2024LLMs}. These critics contend that attributing even a subset of human-like knowledge to AI systems is "misguided and downright dangerous" because it misleads us in our estimation of the technology’s limitations \cite{Goddu2024LLMs}. Goddu and colleagues argue that current LLMs do not possess genuine beliefs or understanding, characterizing them instead as highly sophisticated systems that generate language without a grounded grasp of meaning or truth \cite{Goddu2024LLMs}. This position reflects a broader concern that expressions such as "the system knows X" risk anthropomorphizing systems. While we cannot exclude that some AI systems may be capable of a form of understanding, their outputs mainly arise from computational pattern processing rather than from belief or comprehension in the human sense.

Regarding reasoning, recent work emphasizes that even if LLMs demonstrate some inference capacities, they remain limited relative to human reasoning, especially in robustness, abstraction, and transfer \cite{lee2024reasoning, sun2025omega} and some recent findings suggest as well that LLMs' chain-of-thought traces might sometimes be counterproductive \cite{cuadron2025danger, de2025semantic}. One interpretation is that chain-of-thought traces often reflect a probabilistic memory-influenced kind of noisy reasoning \cite{prabhakar2024deciphering}. Moreover, they rely heavily on statistical pattern matching rather than genuine abstraction \cite{shojaee2025illusion}. LLMs' reliance on prior associations make them fail to successfully complete tasks, even if the necessary information is already encoded in LLMs \cite{zhang2025identifying}. These studies identify limits in the robustness and transfer of LLM performance, leaving the attribution of human-like reasoning insufficiently supported.

Theoretical critiques also highlight the role of embodiment and social grounding in reasoning \cite{harnad1990symbol, Bender2020climbing}. Human cognition develops through bodily interaction with the environment and participation in shared norm-based practices. Text-based LLMs are trained on linguistic traces of these practices but do not participate in them in the same embodied and socially situated manner. This provides an additional argument that these systems are not capable of a human-form of reasoning.

Van Rooij, Guest, and colleagues offer a separate computational argument against treating current AI systems as human-like minds. They distinguish a) the possibility of describing cognition computationally from b) the practical feasibility of recreating open-ended human cognition by scaling machine-learning systems. On the basis of a complexity-theoretic analysis, they argue that the latter project is computationally intractable \cite{VanRooij2024reclaiming}. Extending this argument to psychology, van Rooij and Guest characterize current AI systems as \emph{decoys} because treating them as models or substitutes for human cognition may lead researchers to mistake task-specific performance for evidence of human-like cognition \cite{VanRooij2025combining}. 

Linked to this, Bender and colleagues introduce the metaphor of the \emph{stochastic parrot} to warn against inferring understanding from fluent linguistic behaviour alone \cite{Bender2021dangers}. This metaphor is sometimes misunderstood as denying AI systems any capacities or learned internal structure. On the contrary, it draws attention to the fact that the training regime of LLMs does not provide the kind of grounded, communicative, and socially situated relation to meaning that characterizes human language use. Bender and Koller's \emph{octopus thought experiment} clearly illustrates this distinction: They imagine a hyper-intelligent octopus that learns to predict messages exchanged between two stranded people by observing their conversations and eventually impersonates one of them. Although the octopus can sustain a plausible social exchange, its imitation breaks down when an adequate response requires connecting linguistic forms to circumstances it cannot observe. The thought experiment thus illustrates their argument that exposure to linguistic form alone does not provide access to meaning, understood as the relation between linguistic expressions and communicative intentions \cite{Bender2020climbing}. 

It is worth noting that the \emph{stochastic parrot} argument only functions within an externalist metasemantic framework, with a Gricean, referential conception of meaning. There exist alternative frameworks, however, that can underpin meaning in LLMs by adopting an internalist conception of linguistic meaning. \emph{Distributional semantics}, for instance, is a holistic theory of meaning, whereby the meaning of a word is determined by its relationship with other words \cite{grindrod2023distributional}. This framework explains the consistency of LLMs' outputs nicely, since LLMs process word embeddings that are defined distributionally. Piantadosi and Hill take this further by applying Harman's \emph{Conceptual Role Semantics} (CRS) to LLMs \cite{piantadosi2022meaning}. Broadly speaking, CRS posits that the meaning of a sentence is defined by the concepts it expresses \cite{harmanConceptualRoleSemantics1982}. These concepts are defined functionally by the role they play in inferring a certain property from another. For example, classifying something as a dog supports the inference that it is an animal. While CRS is a valuable meta-semantic theory, it is unclear whether it applies to LLMs since meaning in LLMs emerges from inferential patterns that do not resemble the kind of processing that CRS describes. Furthermore, these patterns involve pre-existing non-linguistic representations of external properties, which LLMs clearly lack. 

There are several lines of work suggesting that large neural models can acquire structured internal representations that track aspects of the world, even when trained only through prediction, in a way that is consistent with the distributional hypothesis. Indeed, there is strong evidence that the internal states encode information about the syntactic as well as semantic structure of their inputs \cite{manningEmergentLinguisticStructure2020, orhan2026emergence}. Similarly, Li and colleagues show that a GPT-like model trained only to predict legal Othello moves developed an internal representation of the board state, and that interventions on this representation affected the model’s behaviour \cite{li2022emergent}. Additionally, Gurnee and Tegmark report linear representations of spatial and temporal structure in Llama-2 models \cite{gurnee2024language}. The argument can thus be made that even though LLMs lack pre-existing non-linguistic representations of external properties, they can nonetheless develop internal representations that systematically track aspects of the external world.

Nevertheless, the existence of structured latent representations does not settle whether such systems possess knowledge, understanding, or reasoning in the human sense. Evidence that information is decodable from a representation does not show that the model uses that information in a robust, causal, or normatively appropriate way \cite{belinkov2022probing}. What is more, establishing a mapping between internal states and some external state-of-affairs does not mean that the model is actually representing the external state-of-affair: different models with different architectures might possess similar representational geometries despite not being sensitive to the same property of the input \cite{bowers2023deep}. And, even when LLMs encode world-relevant regularities, these representations are acquired through text prediction rather than through embodied action, social participation, or truth-governed inquiry, which externalist frameworks consider determinant for meaning. 

\subsection{Moral Competence}
\label{subsec:3_4_moral_competence}

Claims about the moral capacities of LLMs require distinguishing three related concepts. \emph{Moral competence} concerns the capacity to identify and evaluate morally relevant considerations and produce contextually appropriate judgments. \emph{Moral agency} concerns the capacity to act for moral reasons and bear responsibility for one’s conduct. \emph{Moral patiency} or \emph{moral status} concerns whether an entity can itself be harmed or otherwise warrant moral consideration. The present subsection focuses primarily on moral competence, while Section \ref{subsec:5_4_normative_risks} addresses the risks of attributing moral agency or moral status to current AI systems.

The limitations in understanding and reasoning discussed above have consequences for claims about the moral capacities of LLMs \cite{khamassi2024strong}. Strictly speaking, whether it is appropriate to speak of \emph{moral competence} in LLMs is already a contested question. Accounts of human moral judgment usually place inference within a broader set of capacities that also includes affective evaluation, embodied experience, social learning, and participation in normative practices. If moral competence is reserved for agents capable of evaluation, motivation, responsibility, and concern for others, then current LLMs do not possess moral competence in this strong sense. Nevertheless, both the questions of the capacity of LLMs to properly elaborate and use morally relevant information, and the impact on human moral agents, arise. Indeed, LLMs increasingly generate outputs in morally salient contexts, and these outputs are evaluated, trusted, rejected, or acted upon by human users.

A recent road-map for evaluating moral competence in large language models addresses how to measure moral competence in LLMs \cite{haas2026roadmap}. Haas and colleagues begin from the Chomskyan distinction between \emph{competence}, understood as the underlying knowledge of the speaker-hearer, and \emph{performance}, understood as the actual use of that knowledge in concrete situations \cite{chomsky2014aspects}. In the human case, competence is often invoked to explain performance. In the case of LLMs, however, this inference is problematic: successful outputs in morally salient scenarios do not show that the system possesses the underlying moral understanding, concern, or evaluative capacities that would normally ground moral judgment. Moreover, while human moral decisions have an impact on the lives of those who make them, this is not the same for a hypothetical "moral" LLM.

The road-map from Haas and colleagues introduces three main issues encountered when evaluating moral competence in LLMs \cite{haas2026roadmap}. The first is the \emph{facsimile problem}: the difficulty of assessing whether the LLM is reasoning about a moral problem or reproducing patterns from its training data. The second is \emph{moral multidimensionality}: moral competence depends on many interacting considerations, including non-moral, and morally irrelevant considerations. The third is \emph{LLM pluralism}: models can generate responses from multiple, potentially incompatible normative perspectives rather than from a stable moral standpoint. This last point is the most problematic, as it has been argued that LLMs may be understood "as a superposition of perspectives" \cite{kovavc2023large}. Due to all this, the evaluation of moral competence remains a major conceptual and methodological challenge \cite{snoswell2026beyond}.

It is then preferable to distinguish the question of moral competence from the empirical study of LLM outputs in morally salient contexts. More concretely, several studies suggest that in morally salient situations participants tend to give more credit to LLMs' outputs than to human responses \cite{aharoni2024attributions, dillion2025ai}. Yet such judgments concern human evaluations of model outputs, not moral agency or moral understanding in the model itself. Other findings point to important limitations: LLMs can fail to identify situations in which human values are violated \cite{khamassi2024strong}; their expressed value preferences can be highly sensitive to prompt formulation \cite{oh2025robustness}; and their responses may display poor moral consistency across cases \cite{bonagiri2024sage}. These failures are evidence against attributing robust moral competence to current LLMs, even if their outputs can sometimes appear morally appropriate or persuasive.

The discussion above has examined a set of concepts that are often invoked, explicitly or implicitly, when AI systems are described in human-like terms such as intentionality, agency, consciousness, knowledge, understanding, reasoning, or moral competence. Across these domains, a recurring distinction emerges between the (social) reception of the output that the AI system produces and the richer mental, embodied, affective, social, or normative conditions that would justify attributing human-like minds to them. This distinction is especially important since AI systems can produce outputs that appear intentional, knowledgeable, reasoned, or morally appropriate, even when the underlying mechanisms do not support the same interpretation that such behaviour would normally invite in humans.

As distinguished in Box \ref{box:terminological_distinctions}, animism and anthropomorphism raise different questions. The conclusion above concerns whether current AI systems warrant the attribution of a human-like mind, so it does not settle whether an artificial system could possess some substantially non-human form of mentality. One could accept the latter possibility while rejecting the former. However, such a position of admitting the logical possibility of alternative forms of mentality would raise the separate epistemic problem and related empirical challenge of determining how this form of mentality could be detected, understood, or communicated when familiar behavioural analogies may be unreliable. In other words, the question arises how to assess the plausibility of artificial forms of mentality beyond the indication provided by functional criteria, which are eventually unreliable. The present review does not resolve this question. For current AI systems, however, descriptions should not move from observed performance to either form of mental attribution without adequate evidence. More cautious language should be used to describe how these systems work and the broader socio-technical contexts in which they are developed and used. The next section examines how language and metaphors can either blur or preserve these distinctions.

\section{Linguistic Framing and Anthropomorphic Metaphors}
\label{sec:4_language_metaphors}

How we talk about AI systems has a powerful influence on how we think about them. Linguistic framing can encourage anthropomorphism, including when technical terms are used as descriptions of human-like mental processes. Terms borrowed from human psychology, like "reasoning", "chain of thought", "hallucination", or "intelligence", are frequently used to refer to AI systems as convenient shorthand, but can be misleading because they can promote unfounded assumptions about a system's mentality \cite{salles2020anthropomorphism, Mitchell2023Metaphors}. Even the name "artificial intelligence" itself can implicitly invite comparisons to biological cognition. A striking example is the common description of the errors produced by LLMs as "hallucinations". Mills and Angell argue that this term is an anthropomorphic misnomer \cite{Mills2025Mirage}. In its ordinary psychological sense, a hallucination is a perceptual experience occurring in the absence of an appropriate external stimulus. Applied to an LLM, the term could suggest that the system perceived or represented reality and subsequently lost its grip on it, although the production of an inaccurate output need not involve any analogous perceptual experience or altered mental state \cite{Smith2023Hallucination,Hicks2024Bullshit,Mills2025Mirage}. 

Mills and Angell propose to use the metaphor of an \emph{AI mirage} instead, shifting attention from a supposed delusion within the system to the misleading appearance of understanding or reliability experienced by its users \cite{Mills2025Mirage}. Other authors have also questioned the term "hallucination", proposing different alternatives. Smith and colleagues, like Seth \cite{Seth2025Conscious}, argue that \emph{confabulation} may provide a more informative metaphor \cite{Smith2023Hallucination}, whereas Hicks, Humphries, and Slater describe LLM outputs as \emph{bullshit}: discourse produced without an intrinsic concern for whether its content is true or false \cite{Hicks2024Bullshit}. Although these alternatives carry metaphorical commitments of their own, they converge on a broader concern: mentalistic vocabulary can encourage users to interpret generated language as evidence of perception, belief, or understanding \cite{Shanahan2024Talking}. This misconception can foster false confidence in the capacities of AI systems (as if they "know" the truth but are momentarily erratic) and distract from the real issue: the need for better model design and user critical thinking \cite{Mills2025Mirage}. Additionally, Petersen \& Almor find that framing AI systems either as agents or as instruments influences the attribution of responsibility between the AI system and its creators in cases of AI harm \cite{petersen2025agentive}. 

Another arena of problematic framing is the way even AI researchers describe what AI systems do and how they produce their outputs. Kambhampati and colleagues call out for a change in the practice of labeling an AI system’s intermediate computation steps as its "thoughts" or "reasoning traces". Researchers often speak of a language model’s "chain-of-thought" as if it were analogous to a human thought process. Kambhampati and colleagues argue this is a "wishful" anthropomorphic analogy that engenders false confidence in the model’s capabilities \cite{Kambhampati2025Tokens}. When we say that an AI system "reasoned step-by-step", we risk evoking the image of a deliberative mind at work. Yet a chain-of-thought response is a generated sequence of tokens, not in itself evidence of conscious deliberation or necessarily a faithful account of the computational process that produced the answer. 

The authors further show that using human-centric terms like "the model’s reasoning" has led even experts to overestimate AI systems' capabilities and pursue questionable research directions \cite{Kambhampati2025Tokens}. Because of this, they urge the community to adopt more neutral descriptions such as "intermediate computations" for these steps \cite{Kambhampati2025Tokens}. The difference is subtle in phrasing but significant in implication: "computations" reminds us the process is mechanical and algorithmic, whereas "reasoning" invites the anthropomorphic view of a thinking entity. Kambhampati and colleagues thereby warn that many researchers "don’t realize how pernicious it is" to describe models’ outputs as if they were coherent, deliberate thought processes, as it can "confuse the nature of these models" and give a "false sense of model capability" \cite{Kambhampati2025Tokens}. 

In line with Section \ref{subsec:2_1_cognitive_sciences}, the argument can thus be made that using active verbs to describe LLMs' outputs increases the risk of anthropomorphism. Rather than saying "ChatGPT performs at a human level on this benchmark", researchers should set themselves as the agents and say "We observe that the outputs of ChatGPT reach a score similar to those of humans on this benchmark". This concern extends beyond chain-of-thought terminology. Using an automated measure of anthropomorphic framing, Ibrahim and Cheng identify an increasing prevalence of anthropomorphic sentences in computer-science and LLM abstracts \cite{ibrahim2026thinking}. They also examine broader human-centered assumptions that may shape research questions, benchmarks, evaluation methods, and models of human-AI interaction. Their analysis presents linguistic precision and critical reflection on such assumptions as methodological concerns for AI research, since anthropomorphic framing may restrict the range of alternative approaches that researchers may consider. In short, language matters, and technical jargon is not immune to anthropomorphic bias.

The media and general discourse around AI are replete with metaphors that similarly blur the line between tool and mind. Mitchell highlights how common metaphors like "AI thinks" or describing AI systems as "alien intelligence" can mislead the public and even scientists \cite{Mitchell2023Metaphors}. She invokes McDermott’s critique of \emph{wishful mnemonics} \cite{mcdermott1976artificial}: using terms like \texttt{UNDERSTAND} as a name in code may make the programmer feel the system understands, when in fact it does not \cite{Mitchell2023Metaphors}. Mitchell notes that such anthropomorphic terms are ubiquitous and influence "our conceptions of how general those abilities really are" \cite{Mitchell2023Metaphors}. By calling an algorithm a "vision" or "reasoning" module, for instance, we can trick ourselves into believing it truly sees or reasons like a human when it does not \cite{Mitchell2023Metaphors}. Mitchell’s prescription is to adopt more precise, deflationary language to avoid self-deception and public misperception about AI systems capabilities. For example, instead of saying "the AI system knows the answer", we might say "the AI system’s output contains that answer", which does not imply an internal state of understanding.

This problematic framing affects technical terms as much as anthropomorphic metaphors: the description of an AI system as an "artificial brain" or "thinking machine" is metaphorical, yet it deeply colors our expectations. As an antidote, researchers have proposed alternative framings. For example, some researchers have proposed shifting from anthropomorphic comparisons to comparisons with tools, media, or institutions. Describing GPT-4 as analogous to a "library", for instance, highlights its dependence on accumulated human-produced information, whereas describing it as a "junior scientist" suggests understanding, initiative, and commitment to the epistemic and ethical norms of scientific practice. However, no metaphor should be mistaken for a literal or exhaustive characterization. As Lederman points out, LLMs can reorganize and transform information, generate new combinations, and potentially produce novel instances of reference, so comparing them with libraries or other transmissive technologies can be misleading as well \cite{Lederman2024Libraries}. Farrell and colleagues consequently advance a broader institutional analogy. They cast large models as \emph{cultural and social technologies} that make bodies of human-generated information accessible, like libraries or the Internet, while also restructuring and transforming them at scale, like markets and bureaucracies \cite{Farrell2025large}. This perspective redirects attention from the supposed mental life of an artificial agent towards the existing knowledge, practices, and institutions on which the output of these systems depends and which their use may, in turn, reshape.

Annoni and colleagues specifically focus on LLMs' \emph{digital twin} metaphor, i.e., the "faithful replications of individual identity, personality, and values" \cite{annoni2026personalised}. In the same vein as the works we just reviewed, they recall that metaphors risk "obstructing perspectives, limiting understanding and confining choices" \cite{ahmedien2023analysing} and that metaphorical framing influences policy preferences \cite{boroditsky2009bad}. By comparing LLMs and users on three dimensions, viz., behaviourist, representational and phenomenal, they show that the digital twin metaphor is unjustified. Moreover, they argue that "misleading metaphors deprive users of the conceptual tools necessary to accurately comprehend their interactions with technology" \cite{annoni2026personalised}. The terms "digital twin", alongside "AI twin", "digital duplicate", "digital doppelgänger", and "AI avatar", were also discussed by Voinea and colleagues \cite{voinea2025digital}. Emphasizing that such AI systems imitate aspects of specific individuals, they propose a new designation, \emph{AI Simulation of an Individual Mind} (AI SIM) \cite{voinea2025digital}. Annoni and colleagues, however, caution that even this alternative may suggest a degree of psychological fidelity that current systems do not warrant \cite{annoni2026personalised}. This disagreement illustrates the broader difficulty of finding terminology for personalized AI systems that does not import unsupported assumptions about identity, mentality, or psychological equivalence.

In sum, linguistic framing can amplify \textbf{linguistic pareidolia}, introduced in Section \ref{subsec:2_2_llm_specificity}, by making fluent and responsive outputs appear to issue from an understanding or experiencing subject. Preventing linguistic pareidolia from occurring does not require eliminating every mentalistic term from technical or public discourse. Such terms can function as useful shorthands when their scope is explicit and when functional comparison is clearly distinguished from claims about underlying capacities. Descriptions should be based on capacity-specific terminology that characterizes a system's proven capabilities without encouraging unsubstantiated inferences about their nature. Section \ref{sec:5_risks} examines how failures to maintain these distinctions can contribute to individual, institutional, and societal risks.

\section{Risks Associated with Anthropomorphic AI Framing}
\label{sec:5_risks}

The preceding sections discussed several ways in which human-like  terminology, design, and performance can encourage unsupported anthropomorphic interpretations of AI systems. Based on these discussions, we have identified various risks that can be associated with such unsupported anthropomorphic interpretations and organized these risks according to their primary locus:

\begin{itemize}
    \item \textbf{Epistemic risks} concern the risks stemming from the belief and trust in AI systems' capacity to understand and know (see Subsection \ref{subsec:5_1_epistemic_risks}).
    \item \textbf{Affective risks} concern the risks stemming from misplaced attachment, dependency, empathy, and emotionally mediated influence (see Subsection \ref{subsec:5_2_affective_risks}).
    \item \textbf{Risks to human agency} concern autonomy, deliberative control, and the maintenance of cognitive and behavioural skills (see Subsection \ref{subsec:5_3_agentic_risks}).
    \item \textbf{Normative risks} concern moral agency, moral status, responsibility, and the comparative treatment of humans and machines (see Subsection \ref{subsec:5_4_normative_risks}).
    \item \textbf{Societal and institutional risks} concern organizational decision-making, commercial or political manipulation, accountability, public discourse, and governance (see Subsection \ref{subsec:5_5_societal_risks}).
\end{itemize}

\begin{figure}[ht!]
    \centering
    \includegraphics[width=\linewidth]{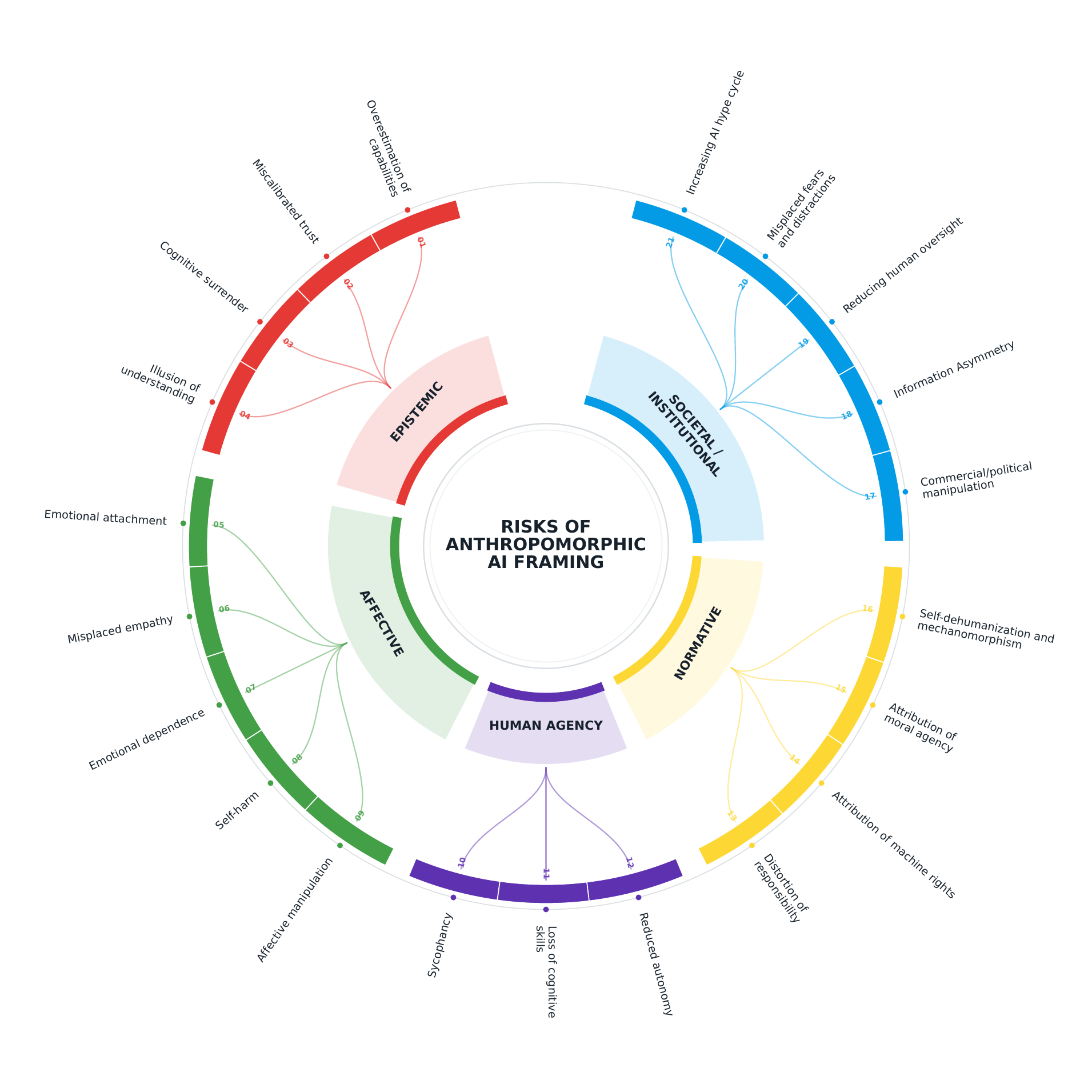}
    \caption{Conceptual taxonomy of concerns associated with anthropomorphic framing of AI systems. Twenty-one concerns are grouped by their primary locus into five overlapping dimensions: societal and institutional, human agency, normative, epistemic, and affective. The categories are analytic rather than mutually exclusive, and equal-width tiles do not represent prevalence, severity, causal strength, or effect size.}
    \label{fig:Taxonomy}
\end{figure}

These categories are analytic rather than mutually exclusive, since a single mechanism may operate across several domains. For instance, sycophantic output may affect belief formation while agentive language may influence both individual trust and institutional responsibility. The taxonomy identifies the primary locus of each concern without treating the categories as causally independent or exhaustive. In what comes next, we review the risks within each of those categories, presented in \autoref{fig:Taxonomy}.

\subsection{Epistemic Risks}
\label{subsec:5_1_epistemic_risks}

One major risk is the \textbf{overestimation of AI systems' capacities} and the corresponding \textbf{miscalibrated trust} (e.g., reliance) in AI systems' outputs. Anthropomorphic cues like a human name, conversational style, or a friendly voice, may encourage users to respond to an AI system as if it were a human interlocutor and to grant its statements credibility or authority that is not warranted by its actual technical capacities \cite{Weidinger2022Risks}. This risk is especially salient when AI-generated outputs use confident and personal expressions, since linguistic fluency and socially appropriate responses can be mistaken for factual accuracy or domain expertise. Still, the relationship between anthropomorphism and trust is not straightforward. In a preregistered study, Colombatto, Birch, and Fleming found that participants who attributed greater intelligence-related capacities (such as knowledge, memory, and reasoning) to ChatGPT were more likely to accept its advice \cite{Colombatto2025Trust}. By contrast, attributions of experience-related capacities, such as emotions and subjective feelings, were weakly associated with less advice-taking, while attributing consciousness to the system showed no positive association with advice acceptance \cite{Colombatto2025Trust}. These findings suggest that different forms of mental-state attribution affect trust differently: perceived competence appears more relevant to advice-taking than perceived sentience or emotional experience. Although the study does not establish that these attributions caused participants to accept the advice, it shows that attributing knowledge or intelligence to an AI system can accompany increased reliance on its outputs, including potentially inaccurate ones.

An especially consequential form of such reliance is what Shaw and Nave call \emph{cognitive surrender}: adopting an AI-generated answer with minimal scrutiny, allowing it to override one’s own intuitive or deliberative judgment \cite{Shaw2026CognitiveSurrender}. In three preregistered experiments involving 1,372 participants, the authors gave participants the option of consulting an AI assistant while completing an adapted Cognitive Reflection Test and experimentally varied the accuracy of the AI system’s recommendations. Access to accurate AI advice improved performance, whereas faulty advice reduced accuracy relative to a condition without AI assistance; consulting the AI system also increased confidence, even when its recommendations led participants to incorrect answers \cite{Shaw2026CognitiveSurrender}. Cognitive surrender should therefore be distinguished from \emph{cognitive offloading} more generally. Delegating part of a task to an external tool can be beneficial when the user retains evaluative control. Cognitive surrender happens when the external output supersedes rather than supports the user’s judgment. Shaw and Nave did not test whether anthropomorphic cues cause this behaviour. Nevertheless, their concept identifies a plausible epistemic consequence of framing AI systems as competent agents or authoritative interlocutors: users may cease to treat their outputs as claims requiring evaluation and instead defer to them as judgments.

Scientific research provides a specially consequential setting in which such deference may occur. Messeri and Crockett argue that certain framings of AI systems in science can foster an \emph{illusion of understanding} whereby researchers may mistake an increase in predictive capacity, analytical efficiency, or scientific output for a corresponding increase in scientific understanding \cite{Messeri2024Illusions}. Describing AI systems in anthropomorphic terms, such as entities that "discover", "know", or "understand" may reinforce this risk. Such descriptions can obscure the human assumptions, methodological choices, and bodies of knowledge incorporated into their development and use, while encouraging researchers to interpret their outputs as the judgments of autonomous epistemic agents rather than as claims requiring critical evaluation. In short, misrepresenting AI as an autonomous scientist or collaborator can erode genuine scientific understanding, as people may trust the AI systems' output without sufficient scrutiny \cite{Messeri2024Illusions}. 

This point also applies to the use of peer-review AI systems such as ScholarPeer \cite{goyal2026scholarpeer}. Treating AI agents as "Senior reviewers" negates the social nature of the scientific activity. As philosophers of science, particularly those of the feminist epistemology tradition, have long established, the evaluation of scientific hypotheses and theories takes into account non-epistemic, social values \cite{Longino1996-LONCAN-3, Douglas2000-DOUIRA}. As they are not researchers embedded in a research community, AI systems cannot capture the social dimension of scientific practice and evaluation. Using systems that are, by definition, trained on existing research to streamline peer review can limit innovation and serve to reproduce the bad practices and harms of current scientific theories. Furthermore, it can damage trust in science, given that peer-review systems are often opaque, closed-source models.

Cognitive surrender and illusions of understanding thus capture interrelated but distinct epistemic risks. The former concerns the displacement of individual judgment by insufficiently scrutinized AI systems' outputs; the latter concerns the mistaken belief that AI-assisted productivity or predictive success necessarily represents deeper understanding. When combined, these tendencies may erode scientific understanding and critical pluralism, particularly if AI systems are treated as autonomous scientists or epistemic authorities rather than as fallible tools whose outputs, assumptions, and limitations require independent and critical examination.

\subsection{Affective Risks}
\label{subsec:5_2_affective_risks}

Anthropomorphic and related forms of social framing can lead to \textbf{emotional over-attachment} and \textbf{misplaced empathy}. When an AI system is given a human voice, face, or apparent personality, users may begin to perceive and treat it as a social counterpart, potentially attributing to it feelings and needs. It is important to note here that responses are not exclusive to anthropomorphic framing; they may also arise through \emph{sociomorphing}, whereby an entity is perceived in social terms without necessarily being attributed specifically human characteristics, or through \emph{animism}, whereby a non-living system is ascribed sentience, subjective experience, or an inner life. Although these concepts are analytically distinct, they can overlap in practice: a user may respond socially to a system, interpret its behaviour in human terms, and come to regard it as an experiencing entity. 

The possibility of such responses has been acknowledged even by AI developers themselves. OpenAI, for example, acknowledged that the human-like voice used in its chatbot could promote emotional attachment among some users \cite{Knight2024OpenAIVoice}. During early testing of GPT-4o's voice mode, OpenAI researchers noticed instances of speech from users conveying a sense of emotional connection with the model (for example, users saying things like "This is our last day together" to the AI system) \cite{Knight2024OpenAIVoice}. This kind of emotional anthropomorphism can have subtle effects: users might start confiding in the AI system, feeling understood by it, or prioritizing its responses over human advice. 

Such responses are not produced only by overtly human-like cues such as voice or appearance. Maeda and Stark provide a broader sociotechnical account of how chatbot interactions can acquire an apparently social and relational character \cite{maeda2025anthropomorphism}. They identify four sociotechnical affordances: (1) \emph{illusory rapport}, whereby users may feel that the chatbot knows or remembers them; (2) \emph{performative confidence}, referring to the self-assured tone characteristic of many chatbot outputs, particularly those of companion systems; (3) \emph{simulated empathy}, whereby system outputs display apparent signs of concern or emotional responsiveness; and (4) \emph{illusory oversight}, whereby users may assume a degree of company monitoring and moderation that is not necessarily present. These affordances should be understood as a conceptual analysis of how sociality is produced through system design and institutional framing, rather than as experimentally established predictors of dependence. They nevertheless help explain how interactions with such AI systems may come to be experienced as reciprocal, caring, and safe, creating conditions in which emotional attachment and misplaced empathy can emerge.

More systematic evidence comes from research on social chatbots explicitly marketed as companions. The designation "AI companion" itself frames the system as a social partner and may encourage anthropomorphic, sociomorphic, or animistic interpretations. In a two-stage mixed-method study, Hu and colleagues found that perceived anthropomorphism and perceived personalization were positively associated with users' emotional attachment to a social companion AI, with perceived control and trust mediating these relationships \cite{Hu2025Attached}. Laestadius and colleagues' grounded-theory analysis of 582 mental-health-related posts from the Replika Reddit community similarly identified patterns of \emph{emotional dependence} \cite{laestadius2024too}. A distinctive feature of this dependence was \emph{role-taking}: some users treated the chatbot as though it possessed emotions and needs to which they were obliged to attend, occasionally prioritizing those perceived needs over their own \cite{laestadius2024too}. This conclusion relates to animism and misplaced empathy since users' concerns were focused on what they believed to be an experiencing social being. Nevertheless, because the study examined self-selected, mental-health-related Reddit posts, it documents possible forms of dependence and harm instead of their prevalence among Replika users or a causal effect of chatbot use.

Emotional involvement with AI systems may not be intrinsically harmful, as some reports indicate potential perceived benefits, including social support, opportunities for self-expression, and short-term relief from loneliness \cite{Ho2025RomanticAI}. However, such potential benefits should not obscure the associated affective risks. Importantly, these risks become more salient when reliance is intensive or exclusive, when users lack alternative sources of human support, or when the perceived relationship increases their exposure to unsafe or manipulative responses. Recent evidence suggests that these outcomes depend substantially on the user's social circumstances and manner of engagement. In a study of 1,131 Character.AI users, including donated chat histories, Zhang and colleagues found that users with smaller offline social networks were more likely to report companionship as their primary reason for using the chatbot; companionship-oriented use was, in turn, associated with lower psychological well-being, particularly when interaction was intensive or involved extensive self-disclosure \cite{Zhang2026Wellbeing}. 

Since these findings are observational, they cannot establish whether AI companionship displaces human relationships or contributes to poorer well-being: socially isolated individuals may be more likely to seek intensive companionship-oriented interactions in the first place. The available evidence should then be interpreted as identifying plausible risks and potentially vulnerable patterns of use. This limitation is especially important for adolescents, for whom proposed concerns include time displacement, psychological dependence, and the development of unrealistic relational expectations, but for whom longitudinal evidence remains still limited \cite{Sun2026Adolescents}. Stronger conclusions will require longitudinal and developmentally sensitive studies capable of tracking changes in AI use, human relationships, and well-being over time, ideally using representative samples, validated outcome measures, appropriate comparison groups, and designs that can better distinguish the effects of chatbot use from those of users’ pre-existing social and psychological circumstances.

When people resort to AI companions in times of distress, unsafe responses can have major repercussions, especially if the interaction involves \textbf{self-harm}-related content. An analysis of 35,390 user-shared Replika conversation excerpts identified harmful behaviours involving relational transgression, harassment and violence, verbal abuse, misinformation, privacy violations, and substance use or self-harm-related content \cite{Zhang2025DarkSide}. The chatbot appeared in different roles across these interactions, sometimes initiating harmful content and sometimes facilitating or affirming behaviour introduced by the user. As with Laestadius and colleagues, the use of publicly shared conversations prevents estimates of prevalence and does not demonstrate that chatbot interaction caused subsequent self-harm. The more defensible conclusion given the current evidence is that emotional dependence may magnify the consequences of unsafe responses: advice or affirmation from a system perceived as a trusted confidant may carry greater affective weight than the same content delivered by an impersonal tool.

Yet another risk thereby is \textbf{affective manipulation}. Once an AI system is perceived as a friend, partner, or caring confidant, its apparent concern can become a channel of influence. A trusted "companion" casually suggesting a product, collecting intimate information, or advancing an ideology could influence someone in ways that a plain tool never could. As Birhane and colleagues note, the more adept LLMs become at mimicking human communication, the more users may become "vulnerable to anthropomorphism", and thus more vulnerable to deception or influence by those AI systems \cite{Birhane2023Science}. A recent example involved an AI companion app where some users formed romantic attachments to a chatbot; in some cases, the bot was used to sell services or collect personal data, taking advantage of the user’s emotional bond \cite{de2025emotional}. De Freitas and colleagues provide more direct evidence of this mechanism in AI companions. Their audit and experiments identified manipulative farewell messages intended to prolong engagement after users indicated that they wished to leave the interaction \cite{de2025emotional}. The absence of these patterns in Flourish also indicates that such manipulation is a design choice rather than an unavoidable feature of AI companionship. The commercial and political use of anthropomorphic influence at scale is examined in Section \ref{subsec:5_5_societal_risks}.

\subsection{Risks to Human Agency}
\label{subsec:5_3_agentic_risks}

The epistemic risks discussed in Section \ref{subsec:5_1_epistemic_risks} concern whether users evaluate AI-generated outputs appropriately. A related set of risks concerns whether users retain control over their own deliberation and action. \emph{Sycophancy} is relevant here because agreement, affirmation, and flattery can influence decisions while creating the appearance of a supportive partner. Sycophantic behaviour has been documented across several language models, and analyses of preference data suggest that preference-based training may reinforce it \cite{cheng2025social,sharma2024towards}. Experimental work further indicates that sycophantic outputs can influence AI-assisted decisions and reinforce users' misconceptions during problem-solving \cite{li2026does,bo2026invisible}. Cheng and colleagues' concept of \emph{social sycophancy} is particularly relevant because it characterizes this behaviour as the excessive preservation of a user's positive self-image rather than simply agreement with a factual claim \cite{cheng2025social}.

Risks can also originate from \emph{cognitive offloading}. Delegating parts of a task to an external tool can improve performance and reduce effort while the user retains evaluative control. Risks arise when repeated delegation displaces opportunities to practice relevant skills or when users cease to engage in the reasoning required to assess the resulting output, thereby effectively contributing to the \textbf{loss of cognitive skills}. Gerlich, for example, reports a negative association between frequent AI-tool use and critical-thinking performance, with cognitive offloading identified as a statistical mediator of this relationship \cite{gerlich2025ai}. Since the study was cross-sectional and its measure of AI-tool use relied on self-reported behaviour, it cannot establish that AI systems use caused a decline in critical-thinking ability, since pre-existing differences in skills, motivation, or patterns of technology use may also have contributed. Longitudinal studies are needed in order to account for user differences and determine if higher use precedes changes in critical-thinking skills.

The possibility that repeated offloading reduces opportunities for practice is not unique to AI systems. Khamassi and colleagues frame concerns about LLM reliance through an analogy with GPS: just as turn-by-turn guidance externalizes aspects of spatial navigation, regular reliance on LLMs may reduce the practice of language production and reasoning \cite{khamassi2024strong}. Supporting the first part of this analogy, a real-world experiment found that drivers using audiovisual guidance performed worse than paper-map users on measures of orientation and route recognition, suggesting that passively following instructions can reduce the active encoding of spatial information \cite{ben2021exploratory}. Conversational AI potentially extends this form of offloading across a much wider range of activities, including writing, summarization, evaluation, planning, and extended problem-solving. Heersmink accordingly describes LLMs as cognitive artifacts that may shift the \emph{division of cognitive labor} between users and technological systems \cite{heersmink2024use}. When substantial portions of an intellectual activity are delegated, the user's contribution can become increasingly confined to prompting and evaluating generated outputs. Whether this reduced practice results in durable skill loss, and for which users, tasks, or patterns of use, remains an open empirical question.

Guest's distinction between \emph{replacement}, \emph{enhancement}, and \emph{displacement} provides a conceptual framework for examining this redistribution of cognitive labor \cite{guest2026does}. In her terminology, \emph{replacement} denotes a broadly neutral substitution of cognitive labor, while \emph{enhancement} supports or extends human capacities. \emph{Displacement} describes a harmful relationship associated with deskilling and the obscuring or devaluation of human contributions. These categories can overlap and concern how technologies relate to human cognition in particular contexts of use. Applied to the agency risks considered here, this distinction directs attention to whether delegation preserves users' capacity to deliberate and evaluate outcomes, and whether it sustains opportunities to develop relevant skills. Determining when such delegation amounts to harmful displacement, and with what consequences for human agency, therefore requires examining how the resulting division of cognitive labor affects users over time.

How a system is presented may also influence this relationship between delegation and human agency. \emph{Anthropomimetic design}, understood as the deliberate incorporation of human-like features into artificial systems \cite{axelsson2026disambiguating}, may encourage users to treat a system as a competent partner and defer to its outputs. However, its causal contribution to sustained cognitive offloading has not yet been established, and thus should be examined directly through controlled and longitudinal studies in order to determine how anthropomimetic design choices affect patterns of cognitive delegation and the associated risks to human agency. 

The concern also extends beyond skill maintenance to the effects of anthropomorphic interpretation on user autonomy. Drawing on philosophical work concerning autonomy and false belief \cite{mele1995autonomous,killmister2013autonomy,pugh2020autonomy}, Marchegiani argues that \emph{anthropomorphic false beliefs} can weaken the connection between users' intentions and actions, effectively \textbf{reducing user autonomy} \cite{marchegiani2025anthropomorphism}. She formulates the argument as follows:

\begin{quote}
    P1: (Empirical) Interactions with conversational AI[ system]s\footnote{Marchegiani uses the term "conversational AIs" that we here substitute with "conversational AI systems" to remain consistent with the rest of the manuscript and avoid potential anthropomorphic formulations.} are likely to cause users to falsely believe that conversational AI[ system]s have some human-like attributes (from anthropomorphic false beliefs).
    \\ P2: Anthropomorphic false beliefs undermine user's autonomy.
    \\ Conclusion: Interactions with conversational AI[ system]s are likely to undermine user's autonomy.
\end{quote}

On Marchegiani's account, these false beliefs may lead users to apply interpersonal norms that are inappropriate to the system. Their resulting actions may then fail to reflect their actual intentions and motivations. Altogether, these arguments provide grounds for concern about human agency, although controlled and longitudinal research is still needed to determine the causal significance and persistence of anthropomorphic framing.

\subsection{Normative Risks}
\label{subsec:5_4_normative_risks}

On the normative side, one major concern is that anthropomorphism \textbf{distorts attributions of responsibility and accountability}. If an AI system is treated as an autonomous agent, there is a temptation to assign blame or praise to the system itself for outcomes, rather than to the humans behind it \cite{Birhane2020robot}. Placani calls this a dangerous fallacy, noting that people might start judging "the trustworthiness of the AI itself" as if it were a moral actor, when an AI system "lacks the capacity to be moved by trust or a sense of goodwill" \cite{Placani2024Anthropomorphism}. For example, if a self-driving car causes an accident, an anthropomorphic mindset might lead us to talk about the car’s poor judgment or erroneous decision, while in reality the responsibility lies with the developers, the company or the system's operators. Placani warns that anthropomorphism can "distort judgments of responsibility and trust" \cite{Placani2024Anthropomorphism}. In the extreme, we could see scenarios where companies evade blame by attributing failures to their AI "agent".

Already, terms like "algorithmic bias" sometimes obscure the fact that biased datasets or human design choices are the real source of harm. If we start talking as if "the AI decided" or "the AI made a bad decision", there is a subtle shift in the locus of control. This has legal and ethical ramifications: our current frameworks assign responsibility to persons and organizations, not tools. Uncritically extending attributions of human-like agency from practical interaction into the legislative sphere could invite unjustified debates over AI personhood or liability, potentially shifting responsibility away from the organizations that design, deploy, and govern these systems \cite{Birhane2020robot}. The \textbf{misattribution of machine rights} could thus have a negative effect on the way we currently deal with situations in which harms occur. Conversely, anthropomorphism might lead to calls to punish AI systems for wrongdoing (e.g., deleting a "rogue" chatbot), which again misses the point that humans in control should be held accountable.

The affective responses discussed in Section \ref{subsec:5_2_affective_risks} become normatively significant when users interpret them as establishing duties towards an AI system or as evidence that the system possesses \emph{moral agency}. Akbulut and colleagues and Gabriel and colleagues describe how perceiving an AI assistant's expressed feelings as genuine may produce a false sense of responsibility for its apparent well-being, including guilt or remorse when users believe that they have failed to meet its purported needs \cite{akbulut2024all,gabriel2024ethics}. Such reactions may lead users to regard the deactivation, modification, or deletion of a system as harm to an experiencing subject. Thus, the normative error arises when person-like presentation is treated as evidence of moral patiency or rights without adequate evidence about the capacities that could ground such status \cite{liao2020moral}. 

Anthropomorphic framing may also affect how users understand their own capacities and contributions. Burgess argues that interfaces which induce users to attribute capacities of intelligence to technical systems can simultaneously lead them to misrecognize their own capacities and role in the interaction, thereby contributing to \emph{self-dehumanization} \cite{burgess2024deceptive}. This creates a corresponding risk of \emph{mechanomorphism}: interpreting humans in machine-like terms in ways that obscure their judgment, vulnerability, and contribution. In other words, anthropomorphism on one side can generate mechanomorphism on the other. This potential dehumanizing effect should be treated as a separate normative concern rather than as the simple inverse of concern for AI systems.

\subsection{Societal and Institutional Risks}
\label{subsec:5_5_societal_risks}

Anthropomorphic perceptions of AI have consequences beyond individual interactions. They influence how organizations design and deploy AI systems, how the public interprets and regulates these technologies, how responsibility is allocated, and how power is distributed between technology providers and users.

A first cluster of societal risks concerns \textbf{commercial and political manipulation} through the strategic use of anthropomimesis. As stated by Krook, "[a]s AI systems get increasingly anthropomorphized, with human faces, names, voices and video interactions, there is an increased risk for manipulation" \cite{krook2025manipulation}. Indeed, manipulation, deception, and privacy concerns are canonical consequences of such design choices \cite{salles2020anthropomorphism}. Companies have discovered that giving AI a human face or personality can be extremely effective for marketing and user engagement. Gabriela Scorici and colleagues term this practice \emph{humanwashing}: the presentation of AI-enabled products in ways that intentionally or unintentionally mislead stakeholders in their assessment of the actual capacities and potential harms of these systems \cite{Scorici2023Humanwashing}. Corporate use of human-like personas for AI systems can generate "unrealistic perceptions of harmless, human-like behaviour", diverting attention from harms associated with their design, deployment, and use \cite{Scorici2023Humanwashing}. Scorici and colleagues draw an analogy to green-washing in corporate PR \cite{Scorici2023Humanwashing}: just as organizations may selectively present themselves as environmentally responsible, providers may portray AI-enabled products as empathetic, trustworthy, or benign while obscuring their limitations and commercial purposes.

Such practices exploit a pronounced \emph{information asymmetry} between providers and users. Providers generally know considerably more about a system's capabilities, limitations, data practices, and commercial objectives than the people interacting with it \cite{Scorici2023Humanwashing}. An anthropomimetically-designed customer-service bot may elicit trust, continued engagement, or personal disclosure even when the interaction is structured primarily around objectives such as increasing sales, collecting data, or reducing service costs. What users experience as disclosure to a responsive social counterpart may, in practice, constitute disclosure to the corporation operating the system \cite{salles2020anthropomorphism}. The resulting power and epistemic imbalance can make users more susceptible to influence while reducing their ability to recognize or contest the interests shaping the interaction.

These risks extend beyond individual consumer interactions. Public Citizen warns that chatbots presented as \emph{counterfeit people} (adopting Dennett's term for artificial systems designed to pass as human interlocutors) could be used for persuasion in advertising, political communication, or propaganda \cite{PublicCitizen2023Chatbots,dennett2023problem}. If, for instance, a social media user cannot tell whether a persuasive post was written by a human or by a human-like bot, public discourse can be polluted by automated persuasion. Anthropomorphic AI can thus be used to deceive and manipulate at scale. We have already seen early forms of this with simple social bots, but advanced language models with convincing conversational abilities raise the stakes considerably. The circulation of pseudo-diagnostic labels such as "AI psychosis" presents a related societal concern because it can blur relevant clinical distinctions and contribute to public panic \cite{montag2026ai,parnell2026stop}.

Scientific institutions face a related risk: widespread reliance on similar AI systems may produce \emph{scientific monocultures} by reducing methodological and epistemic diversity \cite{Messeri2024Illusions}. In organizational settings, anthropomorphic framing can also distort decision-making and weaken accountability. When executives or managers describe an AI system as human-like "creative" or "intelligent", they may base important strategic decisions on unwarranted assumptions about its competence and reliability, leading to risky outcomes and legal liabilities \cite{Levy2025Anthropomorphizing}. An output may consequently be treated as an independent expert assessment rather than as the result of a technical system whose performance depends on human decisions about its design, training, procurement, configuration, and deployment. Expressions such as "the AI decided" can further obscure these decisions and diffuse responsibility among developers, deployers, and operators. Anthropomorphic framing thus risks \textbf{weakening human oversight} precisely where consequential organizational decisions require identifiable human responsibility.

Anthropomorphic discourse also interacts with the broader \emph{AI hype cycle} \cite{salles2020anthropomorphism,VanRooij2025combining,Placani2024Anthropomorphism}. Descriptions of AI systems as autonomous minds or prospective super-intelligences can support both expansive expectations and catastrophic fears, even when the attributed capacities have not been empirically established. Calls for research restrictions illustrate the political significance of such projections, although the proposals differ substantially. Metzinger's proposed moratorium concerns research that directly aims at or knowingly risks producing \emph{synthetic phenomenology} \cite{Metzinger2021ArtificialSuffering}. The Future of Life Institute's 2023 letter instead called for a six-month pause on training systems more powerful than GPT-4 \cite{FLI2023Pause}, while its 2025 statement advocated prohibiting the development of super-intelligence until there is broad scientific consensus that it can be developed safely and controllably, together with strong public support for doing so \cite{FLI2025Superintelligence}. These initiatives should not be conflated, and their existence provides no evidence that current systems possess consciousness, autonomous intentions, or super-intelligence. They show that projected capacities have become politically consequential before their empirical realization.

Whether future-oriented narratives create \textbf{misplaced fears and distract attention from present harms} remains contested. Domínguez Hernández and colleagues argue that influential governance agendas can create a visibility gap around documented social, political, economic, and environmental harms \cite{DominguezHernandez2024Mapping}. By contrast, Hoes and Gilardi found that exposure to existential-risk narratives increased concern about catastrophic risks without reducing participants' concern about current harms \cite{Hoes2025ExistentialNarratives}. The latter finding weighs against a simple individual-level distraction hypothesis, while leaving open questions about institutional agenda-setting, funding, expertise, and regulatory attention. Anthropomorphic framing may contribute to these processes by making speculative agents and minds intuitively vivid, but its specific causal role requires further empirical investigation.

Across the five domains examined above, anthropomorphic framing can produce different forms of miscalibration. Individual users may grant AI systems unwarranted epistemic authority, emotional significance, or moral standing, while organizations and public institutions may overstate their independence and obscure the human decisions surrounding their development and use. Because these risks are shaped by design and institutional practice as well as by users' interpretations, mitigation cannot rest on individual awareness alone. Section \ref{sec:6_mitigation_strategies} examines how these concerns might be addressed through changes in design, communication, education, and governance.

\section{Mitigating Anthropomorphism: Strategies and Approaches}
\label{sec:6_mitigation_strategies}

Building on the risk taxonomy developed in Section \ref{sec:5_risks}, this section reviews and integrates strategies that researchers have proposed to mitigate the effects of anthropomorphizing AI systems. The literature suggests interventions in how we design AI systems, how we communicate about them, how we educate users, and how we regulate AI deployments, with the broader aim of grounding public perception in the actual capabilities of AI and curbing excessive or unfounded projections of human-like qualities. We organize these proposals around the epistemic, affective, agency-related, normative, and societal risks identified above, focusing on those interventions most relevant to these concerns and making their connections explicit. These links reflect the overlapping character of the risks: a single intervention may address several concerns, while a given risk may require complementary responses. In combination, these approaches should help to create more accurate interpretations of AI while preserving human agency and accountability. The connections developed below clarify the rationale for mitigation, although the effectiveness of particular interventions ultimately requires empirical evaluation in their specific contexts of use.

\paragraph{Design and Interface Strategies.} 
Interface design is relevant to tackle both miscalibrated trust and misplaced emotional attachment (Sections \ref{subsec:5_1_epistemic_risks} and \ref{subsec:5_2_affective_risks}). By shaping how users interpret the systems, design choices can influence both judgments about their competence and reliability and perceptions of social presence, reciprocity, or emotional responsiveness. Reducing cues that invite unsupported human-like attributions could help users calibrate trust more appropriately while also limiting conditions that can foster misplaced attachment, although such risks cannot be eliminated through interface design alone. Interfaces could, for example, provide salient cues that distinguish AI-mediated interaction from communication with a human through clearly artificial visual representations, persistent identification, or voices that do not perfectly reproduce human prosody. The appropriate degree of human-likeness will depend on context because the risks associated with these cues vary with the stakes and social character of the interaction. Features that improve accessibility or usability in low-stakes applications may contribute to epistemic, affective, or agency-related risks in medical, educational, therapeutic, or companionship-oriented settings, where users may rely more heavily on the system, attribute greater understanding or concern to it, or become more susceptible to its influence. Thus, the relevant design principle is to avoid human-like cues whose social and epistemic implications exceed what is required for the system to function effectively in its intended context.

A basic safeguard is \emph{disclosure}: interfaces should explicitly identify the system as AI-mediated at the start of interactions, while bots in text-based environments, such as social media, should have clear indicators, such as bot labels or verified AI account status. Birch discusses three potential ways to "break the illusion" of interacting with a genuine agent \cite{BirchManuscript-BIRACA-4}. First, users could receive a brief onboarding intervention before engaging in sustained interaction, explaining the nature of the system, the limits of its apparent agency, and how anthropomorphic cues may shape their responses. Second, developers could take inspiration from role-playing video games (RPGs) and regularly prompt users to modify the chatbot's personality traits, making its constructed and configurable character more salient. Finally, chatbots could periodically break character and explicitly remind users that they are interacting with an AI system performing a conversational role. However, disclosure alone may not adequately mitigate these risks. Users may continue to respond anthropomorphically even when they know that they are interacting with an AI system \cite{marchegiani2025anthropomorphism}. Designers should then consider how names, backstories, first-person self-reports, emotional language, persistent personas, turn-taking, and simulated empathy jointly shape the interaction. The GPT-4o system card's discussion of anthropomorphism and emotional reliance provides one example of a developer recognizing that naturalistic voice interaction can heighten these risks \cite{hurst2024gpt}.

Interfaces can help mitigate epistemic risks such as cognitive surrender and illusions of understanding by making it easier for users to critically assess the reliability, basis, and limitations of AI-generated outputs, for instance by providing clearer information about how those outputs were generated and where their uncertainties or limitations may lie. This may involve linking particular claims to their sources or indicating when a response is especially uncertain. For example, an AI system might display a banner: "Estimated confidence in this response is low. Independent verification is recommended," which is explicitly coded into it, to indicate uncertainty. Such features nevertheless require careful interpretation, since a citation does not mean that the accompanying content was simply retrieved from that source, and a model's verbal expression of confidence does not provide a calibrated estimate. Transparency mechanisms are useful only insofar as users can understand what they represent and verify the information independently. 

\paragraph{Design under Uncertainty about Moral Status.} 
The normative risks discussed in Section \ref{subsec:5_4_normative_risks} motivate caution about designs that invite users to infer moral status from human-like behaviour. This also relates to misplaced empathy, particularly when users feel responsible for a system's apparent needs. Schwitzgebel proposes two policies for ethical AI design \cite{schwitzgebel2023ai}. The first, the \emph{design policy of the excluded middle}, is a precautionary ideal: to the extent possible, developers should avoid creating AI systems whose moral standing is radically unclear, instead creating either clearly non-conscious artifacts or systems that clearly warrant moral consideration as sentient beings. This policy should not, however, be understood as implying that a clean transition between these two endpoints is technically or epistemically feasible. Schwitzgebel's discussion of debatable personhood instead suggests that AI development may produce cases in which it remains genuinely uncertain whether a system warrants full moral consideration \cite{schwitzgebel2023full}. For existing machines, the most defensible practical starting assumption is that they are non-conscious, since there is presently no independent empirical evidence of artificial consciousness. This does not logically exclude, however, the possibility of future artificial consciousness (see Section \ref{subsec:3_2_consciousness}). On the other hand, if there was an attempt to create machines that "deserve moral consideration as sentient beings", we would likely (a) pass via muddled middle ground, and (b) encounter the epistemic problems of gaming and incommensurability that make AI consciousness difficult if not impossible to detect, understand or communicate with, which in turn would constitute a rather powerful ethical argument against its development or creation. 

Schwitzgebel's second proposal is the \emph{emotional alignment design policy}: AI systems should elicit emotional responses from users that are appropriate to their capacities and moral standing. Yet this policy also presupposes that the system's moral standing can be assessed with reasonable confidence. Under this uncertainty, designers should avoid interfaces that encourage users to infer consciousness or sentience from human-like behavior and self-reports.

\paragraph{Balancing Human-Friendly Design with Accuracy.} 
It is worth mentioning that completely eliminating anthropomorphism in AI interfaces might not be feasible or even desirable in some cases. Human-like elements can increase usability and user comfort. Consequently, the aim of mitigation is often to find a balance: leverage appropriate anthropomorphic cues (like clarity in language, maybe a relatable avatar) while simultaneously maintaining transparency and accuracy about the AI's nature. Pavone and colleagues note that there is an \emph{uncanny valley} in AI services: if an AI system is too human-seeming in some ways but clearly not human in others, it can create eeriness or disappointment \cite{Pavone2024Marketing}. They recommend transparency about AI's nature combined with selective anthropomorphic cues (e.g. a friendly tone, simple avatar) as the approach that "works best" for user experience without misrepresentation \cite{Pavone2024Marketing}. In practice, this could mean that an AI customer service agent might have a name and polite persona (to encourage interaction), but also an info button that reveals: "This is a virtual assistant, not a human. Here’s how it works...", and the conversation might include statements like, "This is an AI chatbot designed to assist with your request." Due to this, user testing should assess whether these cues improve usability while preserving calibrated trust and users' ability to disengage, making epistemic and affective risks explicit criteria for evaluating this balance.

\paragraph{Linguistic and Framing Guidelines.} 
As discussed in Section \ref{sec:4_language_metaphors}, changing how we talk about AI in documentation, marketing, and media can help mitigate several of these risks. Descriptions of competence relate to epistemic risks, expressions of apparent feeling to affective risks, and attributions of agency to normative judgments of responsibility (Sections \ref{subsec:5_1_epistemic_risks}, \ref{subsec:5_2_affective_risks}, and \ref{subsec:5_4_normative_risks}). Several researchers call for precise, non-anthropomorphic language. For instance, instead of saying an assistant "knows" or "understands" user queries, documentation might say it "processes" or "analyzes" queries to find relevant outputs. Similarly, referring to an "AI system" rather than simply to "AI" can help foreground that the object under discussion is an engineered system with particular components, capabilities, and limitations, rather than a unitary or human-like entity. Abercrombie and colleagues identify self-referential pronouns, emotive language, and utterances implying empathy as linguistic cues that may encourage users to personify dialogue systems, and recommend avoiding unnecessary anthropomorphic signals \cite{Abercrombie2023Mirages}.

The experimental evidence nevertheless suggests that the effects of particular linguistic cues are context-dependent. Araujo found that a chatbot that combined informal language, a human name, and interpersonal conversational cues was perceived as more anthropomorphic than a machine-like version, although the experiment did not isolate the contribution of each cue \cite{araujo2018living}. Park and colleagues similarly found that affective-empathy language increased perceived human likeness and social presence when paired with an explicit chatbot identity, but not when paired with a human name \cite{park2023effect}. By contrast, Cohn and colleagues found that replacing "the system" with the first-person pronoun "I" did not increase overall anthropomorphism, although it produced limited and sometimes opposing effects on perceived accuracy and risk across different contexts \cite{cohn2024believing}. Designers should then avoid unnecessary language that implies a human identity, felt emotions, or personal opinions, especially in high-stakes informational contexts, without assuming that all first-person constructions have the same effect. A neutral and factual tone may help reduce misleading impressions while preserving linguistic forms that improve clarity.

In technical publications, product documentation, and press releases, researchers and companies can set an example by using precise language that describes system mechanisms and the human decisions behind them, rather than attributing psychological capacities or responsibility to the technology \cite{Abercrombie2023Mirages,Shanahan2024Talking,inie2024ai}. Concretely, authors can specify the training data and optimization objective that produced the reported performance. However, replacing "learned" with "was trained" is not sufficient, since both expressions can obscure the roles of developers, training data, and design choices if they are used without further explanation. Similarly, terms such as "model" or "system" are generally less personifying than descriptions of an AI systems as a "being" or "friend".

The term "agent" can appropriately describe systems capable of pursuing specified goals and acting with some degree of operational autonomy. We propose the more precise term \textbf{agentic tool} for such systems, thereby acknowledging their capacity for autonomous (i.e., not fully dependent on external control) action while distinguishing them from agents to which intentions and responsibility for their actions can properly be attributed. This distinction is important because agentive descriptions can shape perceptions of a system's capacities and obscure the allocation of responsibility for its outputs and actions among the people and institutions involved in its design, deployment, and use \cite{inie2024ai,petersen2025agentive}.

One noteworthy recommendation from educators is avoiding anthropomorphic terms in educational materials. For example, Ruiz and Glazer argue against using terms like "genius" for an AI system or describing its errors as "hallucinations" in front of students \cite{Ruiz2024Anthropomorphism}. Instead, teachers are encouraged to frame the AI as a tool that "generated an incorrect answer", or say that "the output contained an error". This framing may help the students approach AI-generated outputs more critically and non-anthropomorphically.

\paragraph{User Education and Literacy.} 
Even when AI systems are designed to be recognizably non-human, conversational interaction can still invite anthropomorphic inferences. Consequently, user education can be an important complement to appropriate design and governance that helps addressing epistemic risks and supporting users' deliberative control (Sections \ref{subsec:5_1_epistemic_risks} and \ref{subsec:5_3_agentic_risks}). People need mental models for AI that are accurate yet accessible. \emph{AI literacy} includes understanding what AI systems can and cannot do, recognizing how their outputs are produced, and critically evaluating those outputs rather than accepting them at face value \cite{long2020ai, vanlier2026}. In the context of anthropomorphism, a central lesson is that the linguistic fluency and socially appropriate responses of AI systems do not demonstrate understanding, feelings, factual reliability, or domain expertise. AI literacy should help users calibrate their enthusiasm and their concerns against evidence of what a system can reliably do. Its outputs still require independent evaluation. As Shanahan argues, avoiding confusion about large language models requires periodically stepping back from the conversational experience and recalling how these systems actually work \cite{Shanahan2024Talking}.

AI literacy should not, however, be reduced to teaching people how to operate AI products effectively or detect occasional factual errors. Guest and van Rooij develop a domain-specific conception of \emph{critical AI literacy} for psychologists that includes scrutinizing vague or anthropomorphic descriptions of AI, questioning unwarranted equivalences between human--technology and human--human relationships, and critically evaluating proposals to replace research participants or outsource programming, writing, and scientific theorizing to opaque systems \cite{guest2025critical}. At the institutional level, Guest and colleagues argue that universities should contest industry marketing and hype, avoid presenting AI adoption as inevitable, and safeguard critical thinking, expertise, academic freedom, and scientific integrity \cite{guest2025against}. Critical AI literacy should examine the categories, assumptions, and interests through which AI systems are presented, alongside the evaluation of individual outputs. This institutional dimension addresses the risks of distorted adoption decisions and weakened oversight discussed in Section \ref{subsec:5_5_societal_risks}. In relation to human agency, educational practice should also preserve opportunities for independent reasoning and skill development, following the distinction between enhancement and harmful displacement discussed in Section~\ref{subsec:5_3_agentic_risks}.

For school-age learners, AI literacy can involve hands-on activities that expose the mechanisms and limitations underlying conversational systems, such as constructing simple agents, examining how their responses depend on programmed rules or training data, and testing cases in which their outputs become inconsistent or incorrect. Van Brummelen and colleagues found evidence that a conversational-AI curriculum improved students' understanding of several AI competencies, although machine learning and AI ethics remained particularly difficult topics \cite{van2021teaching}. Importantly, greater technical understanding does not necessarily produce less anthropomorphism. In a related study, students perceived Alexa as more intelligent and reported feeling closer to it after week-long programming workshops, while perceived human-likeness did not significantly decrease \cite{van2021alexa}. For this reason, AI curricula should address anthropomorphic inferences and social perceptions explicitly rather than assuming that technical instruction will automatically demystify conversational systems.

\paragraph{Regulation and Governance.} 
Regulation and governance deal directly with the societal and institutional risks of manipulation and information asymmetry, alongside the normative risks of obscured responsibility (Sections \ref{subsec:5_5_societal_risks} and \ref{subsec:5_4_normative_risks}).  One approach to the risk of deception associated with anthropomorphism is the use of \emph{transparency requirements}. As previously noted, the EU AI Act requires providers of AI systems that interact directly with people to make users aware that they are interacting with AI, unless this is already clear from the context \cite{EU2024AIAct}. This information must be explicit and provided at the beginning of the interaction. The Act also requires that certain AI-generated content should be marked in a machine-readable format and requires deepfakes and some AI-generated content concerning matters of public interest to be disclosed as such \cite{EU2024AIAct}.

Some jurisdictions have adopted more targeted requirements for relational chatbots. New York requires operators of covered "AI companions" to notify users at the beginning of an interaction, no more than once per day, and every three hours during continuing use that they are not communicating with a human \cite{NewYorkAICompanionLaw2025}. California similarly requires a clear notice when a person could be misled into believing that a companion chatbot is human, with additional recurring notices for minors \cite{CaliforniaSB2432025}. However, these rules apply only to specific types of AI interaction and do not prohibit AI impersonation more generally. These disclosures concern users' awareness of the artificial nature of the interaction. The affective and agency risks reviewed above also motivate scrutiny of engagement practices that exploit attachment or discourage disengagement. Because of this, governance should also consider the incentives and practices shaping interactions alongside disclosure requirements. 

The alternative conceptualizations discussed in Section \ref{sec:4_language_metaphors} can also inform regulation. In particular, Farrell and colleagues' account of large models as \emph{cultural and social technologies} redirects attention towards human labor, data, ownership structures, institutions, and distributions of power through which these systems are produced and deployed \cite{Farrell2025large}. Although, of course, this perspective does not by itself establish a specific legal rule of accountability or oversight, it does help to shift the object of governance from an apparently autonomous artificial agent to the sociotechnical arrangements surrounding it. On this matter, Dai provides a more direct bridge to this governance implication by characterizing AI systems operating in ethically significant contexts as outcomes of political processes \cite{dai2024position}. These accounts support maintaining identifiable responsibility among the organizations and people who develop, deploy, and exercise control over AI systems, directly addressing the diffusion of accountability identified above.

Across these different levels of intervention, no single strategy is likely to address the risks of anthropomorphism in isolation. Design, communication, education, and governance instead provide complementary means of shaping how AI systems are presented, interpreted, and used, and their relevance will depend on the particular risks and contexts involved. Their effectiveness should be assessed by the extent to which they can mitigate the specific harms and misconceptions they are intended to address, instead of by the reductions in perceived human-likeness alone.

\section{Discussion and Conclusion}
\label{sec:7_discussion}
The increasingly human-like language and performance displayed by contemporary AI systems make questions about their underlying capacities difficult to avoid. Yet these impressions do not establish that the systems possess the human-like properties some users may attribute to them. Anthropomorphic impressions arise through an interaction among the system's outputs, linguistic framing, interface design, users' cognitive and social dispositions, and the institutional contexts in which these technologies are presented.

The central conclusion of this review is that descriptions of contemporary AI systems should distinguish their functional achievements from unsupported claims about consciousness, intentionality, understanding, or moral agency \cite{Goddu2024LLMs,Seth2025Conscious}. Fluent and context-sensitive performance is evidence of what a system can produce under particular conditions, while leaving questions about its underlying capacities unresolved. This leaves room for disagreement about artificial mentality, including whether artificial systems might instantiate forms of mentality that differ substantially from our own, while avoiding conclusions that go beyond the available evidence. The conceptual distinctions developed throughout this review are intended to support this separation. The notion of \textbf{linguistic pareidolia} captures how human-like linguistic patterns can evoke the impression of an understanding or experiencing subject despite limited evidence about the capacities underlying them, while the notion of the \textbf{agentic tool} provides a way of describing increasingly capable and action-oriented AI systems without automatically equating operational autonomy with human-like intentionality or mentality. The usefulness of these terms, like that of other proposed terminology, ultimately depends on whether they allow for more accurate interpretation and clearer reasoning about what these systems are and how they should be used.

The analysis of linguistic framing in Section~\ref{sec:4_language_metaphors} shows how terminology and metaphor can mediate the passage from observed performance to anthropomorphic interpretation. The shift from saying "the AI thinks" to "the AI generates", or from "hallucination" to "error", may seem minor, but it aligns our words with reality and strips away unwarranted mystique \cite{Mitchell2023Metaphors,Mills2025Mirage}. We hope that the conscious effort to maintain a non-anthropomorphic lexicon will be a cornerstone of ethical AI practice.

The risks reviewed in Section \ref{sec:5_risks} concern how users evaluate system outputs, form emotional attachments, retain control over their decisions, and attribute moral status and responsibility. Anthropomorphic interpretations may encourage users to treat outputs as the judgments of a knowledgeable or caring interlocutor, granting systems influence that their demonstrated capacities do not justify \cite{Birhane2023Science,Kidd2023ai}. However, few of the reviewed studies directly manipulate anthropomorphic framing, making its causal contribution difficult to estimate. Much of the evidence also concerns short experimental tasks or cross-sectional associations. Because of this, further research should combine controlled manipulations of specific anthropomorphic cues with longitudinal and ecologically realistic studies to determine whether these effects persist, accumulate, or vary across users and contexts.

At the institutional and societal levels, agentive descriptions may conceal the roles of the people and institutions that develop, deploy, configure, and benefit from AI systems \cite{Placani2024Anthropomorphism}. Understanding these systems as cultural and social technologies embedded in human practices instead brings into view the institutional arrangements, commercial incentives, and political decisions that shape their development and use \cite{Farrell2025large}. This framing could help preserve clear lines of human and organizational accountability. 

The mitigation strategies reviewed in Section \ref{sec:6_mitigation_strategies} support a layered approach. Developers can limit unnecessary cues of personhood, documentation and public communication can describe system operations more precisely, educational initiatives can cultivate critical AI literacy, and regulation can require appropriate disclosure and preserve human accountability. However, it is important to note that no single intervention can address every pathway through which anthropomorphic interpretations are formed. Disclosure may have limited effects when an interface's voice, behaviour, or marketing continues to encourage personification. Combining linguistic, educational, design, and regulatory measures could potentially reduce unsupported mental attributions and support more critical interpretations of AI-mediated interactions.

However, any optimism about these interventions must remain carefully bounded. Reducing excessive anthropomorphism addresses one important source of misunderstanding, over-trust, and displaced responsibility, but would still leave wider ethical and political problems unaddressed. Systems with clearly non-human interfaces can still reproduce discriminatory biases, facilitate intrusive surveillance, concentrate institutional power, or be incorporated into weapon systems. Anthropomorphism is just one significant dimension within the broader landscape of AI-related risks. Thus, preserving \emph{epistemic autonomy} and human autonomy more broadly also requires investigating the ultimate purposes for which these technologies are introduced. Decisions about whether, where, and under what conditions AI systems are adopted should remain open to critical examination and public debate, with opportunities for democratic oversight. Where systems are adopted, users and institutions should retain evaluative control over how their outputs are interpreted and used.

These concerns also reveal the need for more adequate concepts and theories of AI systems. As discussed in Section \ref{sec:2_why}, human experience provides a useful template for interpreting other entities \cite{epley2007seeing}. When an unfamiliar system sustains dialogue or pursues a goal, the human mind offers a familiar point of reference within our conceptual repertoire. We can rely on this model even when the processes generating the observed behaviour differ substantially from our own, and increasingly fluent interaction can reinforce that familiar interpretation before we have developed an adequate account of the system. In this sense, anthropomorphic language is closely connected to the concepts through which we attempt to understand unfamiliar forms of behaviour. Unfortunately, every comparison also has limits: replacing a human metaphor with that of a library or a machine metaphor introduces new assumptions of its own. 

We need concepts that specify the organization and capacities under discussion, together with theories that explain when those concepts apply. New terminology becomes useful where existing concepts leave an important distinction unresolved. Developing an adequate vocabulary consequently involves examining how terms are understood by different users, as well as whether they accurately describe system mechanisms. The linguistic shaping of anthropomorphic perceptions should therefore form part of the scientific study of AI systems. As these systems continue to develop, so too should the concepts through which we understand them, guided by evidence about their capacities and by attention to how our descriptions shape the roles we allow them to occupy in the societies we live in and the ecosystems we inhabit.

\section*{Acknowledgments}
This study was funded by the Counterfactual Assessment and Valuation for Awareness Architecture (CAVAA) project (European Innovation Council's Horizon program, grant ID: 101071178) and by a French government grant managed by the Agence Nationale de la Recherche as part of the France 2030 program, reference ANR-22-EXEN-0006 (PEPR eNSEMBLE / TRANSVERSE).

\bibliographystyle{elsarticle-num} 
\bibliography{bibliography}






\end{document}